\documentclass[11pt,a4paper]{article}

\usepackage[utf8]{inputenc}

\pdfoutput=1
\usepackage{jcappub}

\usepackage{bm}
\usepackage{amsfonts}
\usepackage{subfigure}

\renewcommand\({\left(}
\renewcommand\){\right)}

\newcommand{\be}{\begin{equation}}
\newcommand{\ee}{\end{equation}}
\newcommand{\bea}{\begin{eqnarray}}
\newcommand{\eea}{\end{eqnarray}}

\renewcommand{\vec}[1]{\boldsymbol{#1}}

\newcommand{\Fourier}[2]{\mathcal{F}_{#1}\left[ #2 \right]}
\newcommand{\InvFourier}[2]{\mathcal{F}^{-1}_{#1}\left[ #2 \right]}
\newcommand{\CLASS}{\textsc{class}}
\newcommand{\CAMB}{\textsc{camb}}
\newcommand{\zadisp}{\vec{\psi}}
\newcommand{\zadispj}{\psi}
\newcommand{\GaussR}{\mathcal{R}}
\newcommand{\ii}{i}

\usepackage[scr=boondox]{mathalfa}

\usepackage{lmodern}
\usepackage{slantsc}

\newcommand{\CONCEPT}{\textsc{co\textsl{n}cept}}

\usepackage{mathtools}

\DeclareMathOperator{\sinc}{sinc}

\newcommand{\textttj}[1]{\texttt{#1}}

\usepackage{textalpha}

\begin{document}
%%%%%%%%%%%%%%%%%%%%%%%%%%%%%%%%%%%%%%%%%%%%%%%%%%%%%%%%%%%%%%%%%%%%%%
%%%%%%%%%%%%%%%%%%%%%%%%%%%%%%%%%%%%%%%%%%%%%%%%%%%%%%%%%%%%%%%%%%%%%%

\title{\textnu{}CO\textit{N}CEPT: Cosmological neutrino simulations from the non-linear Boltzmann hierarchy}

\author[a]{Jeppe Dakin,}
\author[a,b]{Jacob Brandbyge,}
\author[a]{Steen Hannestad,}
\author[b]{Troels Haugb{\o}lle,}
\author[a,c]{Thomas Tram}

\affiliation[a]{Department of Physics and Astronomy, University of Aarhus, Ny Munkegade 120, DK--8000 Aarhus C, Denmark}
\affiliation[b]{Centre for Star and Planet Formation, Niels Bohr Institute \& Natural History Museum of Denmark, University of Copenhagen, {\O}ster Voldgade 5-7, DK--1350 Copenhagen, Denmark}
\affiliation[c]{Aarhus Institute of Advanced Studies (AIAS), Aarhus University, DK--8000 Aarhus C, Denmark}

\emailAdd{dakin@phys.au.dk, jacobb@phys.au.dk, sth@phys.au.dk, haugboel@nbi.ku.dk, thomas.tram@aias.au.dk}

\abstract{In this paper the non-linear effect of massive neutrinos on cosmological structures is studied in a conceptually new way. We have solved the non-linear continuity and Euler equations for the neutrinos on a grid in real space in $N$-body simulations, and closed the Boltzmann hierarchy at the non-linear Euler equation using the stress and pressure perturbations from linear theory. By comparing with state-of-the art cosmological neutrino simulations, we are able to simulate the non-linear neutrino power spectrum very accurately for small to moderate neutrino masses. This translates into a negligible error in the matter power spectrum, and so our \textnu{}\CONCEPT{} code is ideally suited for extracting the neutrino mass from future high precision non-linear observational probes such as EUCLID.
}

\maketitle

%%%%%%%%%%%%%%%%%%%%%%%%%%%%%%%%%%%%%%%%%%%%%%%%%%%%%%%%%%%%%%%%%%%%%%%%%%%%%%%%%%%%%%%%%%%%%%%
\section{Introduction}

The large scale structure in our Universe has been mapped to unprecedented precision during the past decade and provided a spectacular amount of information on cosmological parameters. Within the coming few years large scale structure surveys such as EUCLID~\cite{EUCLID} and LSST~\cite{LSST} will increase the available amount of data by yet another order of magnitude. These surveys are likely to provide the first evidence for non-zero neutrino masses, and eventually enable a precise measurement of the neutrino mass.
This is possible because neutrinos have a significant influence on the formation of structure and lead to damping of fluctuations on small scales. However, the sensitivity of large scale structure formation to the neutrino mass also requires neutrinos to be modelled accurately in e.g.\ $N$-body simulations. 

In order to follow non-linear structure formation of collision-less species it is necessary to solve the collision-less Boltzmann equation,
\begin{equation}
\frac{\mathrm{d}}{\mathrm{d}t}f({\bm x},{\bm p},t) = 0 \, .
\end{equation}
However, solving the equation in the full 6+1 dimensional case is currently not numerically feasible. For cold dark matter the problem can be greatly simplified because the CDM particles have no thermal velocity, reducing the problem to effectively 3+1 dimensions while the perturbations are linear. The most commonly used method for tracking structure formation with CDM is to represent the distribution function, $f$, with particles and follow these in phase space.

Unfortunately, neutrinos cannot easily be followed in the same way because their thermal velocities are larger than, or comparable to, the gravitationally induced streaming velocities. Several schemes have been devised for studying non-linear neutrino structure formation:
\begin{itemize}
\item Using a particle representation of the full neutrino distribution function (e.g.\ \cite{Brandbyge:2008rv,Viel:2010bn,Agarwal:2010mt,Bird:2011rb,Villaescusa-Navarro:2013pva,Castorina:2015bma,Emberson:2016ecv,Adamek:2017uiq}). This requires a much larger number of particles than for CDM because the momentum dependence of the distribution function must be tracked. Furthermore, if the simulation is started early the neutrino structures will be completely noise dominated because of the large thermal velocities. (However, see~\cite{Banerjee:2018bxy} for a radical approach to reducing this noise.)
\item Assuming that neutrino perturbations remain linear~\cite{Brandbyge:2008js,AliHaimoud:2012vj,Liu:2017now}. A simple scheme which is known to work well for small neutrino masses is to use the linear neutrino density field calculated by realising the linear neutrino transfer function on a grid \cite{Brandbyge:2008js}. An improvement on this is to solve the linear theory neutrino equations, but use the full non-linear gravitational potential calculated in the simulation~\cite{AliHaimoud:2012vj,Liu:2017now}.
However, in both cases this scheme only works for small neutrino masses where neutrino perturbations remain linear at all times. 
\item A hybrid combination of the 2 former methods where the neutrino component is initially followed with linear theory but later on, as the thermal velocities approach the gravitationally induced streaming velocities, followed with $N$-body particles \cite{Brandbyge:2009ce}. The same general idea was recently pursued in~\cite{Bird:2018all} where the linear response approximation was combined with a particle implementation.
\end{itemize}

Here, we want to take a somewhat different approach. We start from the full momentum-dependent Boltzmann equation and use the BBGKY~\cite{bbgky1,bbgky2,bbgky3} approach to turn this into a hierarchy of velocity moment equations. For a perfect fluid this hierarchy closes at order 1 and leads to the continuity and Euler equations. However, because neutrinos have a large anisotropic stress component we need to go beyond order 1 in the hierarchy. We demonstrate that by solving the two first moment equations in full non-linear theory while treating the stress and pressure perturbations in linear theory (scaled by the non-linear density field) leads to very accurate results.

We note that the approach of closing the equations at the second moment was also pursued in \cite{Banerjee:2016zaa}, where the Boltzmann equation for neutrinos is similarly recast into hierarchy form, and the integrated (fluid) equations are then solved. However, in \cite{Banerjee:2016zaa} the solution is restricted to non-relativistic particles and the moment hierarchy is closed using an estimate of the second moment gained from the motion of test particles. In this work we use a version of the second moment which guarantees that the solution has the correct behaviour in the linear regime while also allowing the densities and velocities to become non-linear. Furthermore the method presented here works for both relativistic and non-relativistic fluids.

The paper is structured as follows: In section~\ref{sec:theory} we describe the theoretical considerations needed to formulate the hierarchy equations. In section~\ref{sec:the_linear_computation} we discuss the needed linear theory evolution and how to set up initial conditions for the simulations. Section~\ref{sec:implementation_details} contains a review of the numerical methods employed in the simulations, and section~\ref{sec:results} gives a discussion of our main results. Finally, section~\ref{sec:conclusions} contains our conclusions.

%%%%%%%%%%%%%%%%%%%%%%%%%%%%%%%%%%%%%%%%%%%%%%%%%%%%%%%%%%%%%%%%%%%%%%%%%%%%%%%%%%%%%%%%%%%%%%%
\section{Theory}\label{sec:theory}
\subsection{The non-linear Boltzmann equation}
The Boltzmann equation is an evolution equation for the distribution function (a function of 7 parameters) which we choose in their covariant form, namely $x^\mu$ and $P^i$. This 7-dimensional problem can be recast into a 4-dimensional one by taking moments of the Boltzmann equation. The reduction of the dimensionality from integrating out the momentum dependence comes at the price of an infinite hierarchy of moment equations.

The symmetric energy-momentum tensor has 10 independent components which are related to the distribution function, $f$, via the integration
\begin{equation}
	T^\mu_\nu \equiv \sqrt{-g}\int \mathrm{d}^3P\mspace{1.5mu} f \frac{P_\nu P^\mu}{P_0} \, .
\end{equation}
These 10 components can be used to define 10 fluid variables, namely $\delta$, $u^i$, $\delta P/\delta\rho$ and $\sigma^i_j$. The relations are given by
\begin{align}
	T^0_0 &= - \bar\rho(1+\delta) \, ,\\
	T^i_0 &= -\bar\rho\left(1+\delta+w+ \frac{\delta P}{\delta\rho}\delta\right)u^i \, ,  \\
	\theta & = \partial_i u^i \, , \\
	 T^i_j &= \bar\rho\left(w+\frac{\delta P}{\delta\rho}\delta \right)\delta^i_j +\bar\rho\left(1+\delta+w+ \frac{\delta P}{\delta\rho}\delta\right) \left(u^i u_j + \sigma^i_j \right)  \, ,  \label{eq:EMT}
\end{align}
with $\bar\rho$ the average density, $\bar P$ the average pressure and $w\equiv \bar P/\bar \rho$. Finally, $\delta^i_j$ is the Kronecker delta and $\sigma^i_j$ is traceless. These terms are progressively higher order in the velocity expansion. The zero order term is $\delta \rho$, first order terms are $u^i$ and $\theta$, while $\delta P$ and $\sigma^i_j$ are second order terms.

The moment equations can be found by integrating the non-manifestly covariant Boltzmann equation for $f$ (see \cite{Debbasch1,Debbasch2}),
\begin{equation} 
	P^\mu \frac{\partial f}{\partial x^\mu} - P^\mu P^\lambda \Gamma^i_{\mu\lambda}\frac{\partial f}{\partial P^i} = 0 \, ,
	\label{eq:BoltzmannEq}
\end{equation}
over the invariant volume element $\mathrm{d}^3P^{\mathrm{I}}$, which for contravariant momentum variables $P^i$ is given by 
\begin{equation}
	\mathrm{d}^3P^{\mathrm{I}} = \frac{mc}{P_0} \sqrt{-g} \mspace{1.5mu} \mathrm{d}^3P \, ,
\end{equation}
with $\mathrm{d}^3P = \mathrm{d}P^1\mspace{1.5mu} \mathrm{d}P^2\mspace{1.5mu} \mathrm{d}P^3$.

The following moment equations will be derived in the conformal Newtonian gauge with line-element \cite{Ma:1995ey}
\begin{equation}
	\mathrm{d}s^2 = -a^2(1+2\psi)\mspace{1.5mu}\mathrm{d}\tau^2 + a^2(1-2\phi)\mspace{1.5mu}\mathrm{d}x^2 \, .
\end{equation}

\subsection{The moment equations}
\paragraph{Zeroth moment: The continuity equation}\mbox{}\\
Multiplying the Boltzmann equation~\eqref{eq:BoltzmannEq} with $P_0$ and integrating, we recover the general relativistic continuity equation in the weak-field limit:
\begin{align}
	\dot\delta =& -(1+w)(\theta - 3\dot\phi) -3\frac{\dot{a}}{a}\left(\frac{\delta P}{\delta\rho} - w \right) \delta  \nonumber \\
				& -\theta\delta - u^i \partial_i \delta   \nonumber\\
				 &+3\left(1+\frac{\delta P}{\delta\rho}\right)\dot\phi\delta  - \frac{\delta P}{\delta\rho} \theta\delta - u^i \partial_i \left(\frac{\delta P}{\delta \rho}\delta\right)  \nonumber \\
                                & -(\partial_i \psi - 3\partial_i \phi) \left(1+\delta + w + \frac{\delta P}{\delta \rho} \delta \right) u^i \,,
  \label{eq:continuity1}
\end{align}
where a dot implies differentiation with respect to conformal time.

\paragraph{First moment: The Euler equation}\mbox{}\\
To get the Euler equation, we must multiply the Boltzmann equation with $P_0P^i/P^0$, which after integration gives
\begin{align}
	\dot u^i =& -\left[\frac{\dot a}{a}(1-3w) - \dot\psi - 5\dot\phi\right] u^i - \frac{\left[\dot\delta +\dot w + \partial_\tau\left(\frac{\delta P}{\delta\rho}\delta\right)\right]u^i + \delta^{ij} (1+\delta)\partial_j\psi }{1+\delta + w+\frac{\delta P}{\delta\rho}\delta} \nonumber\\
                          &-\frac{1}{\bar\rho\left(1+\delta + w+\frac{\delta P}{\delta\rho}\delta\right)}\left[  \delta^{ik} (\partial_j + \partial_j \psi -3 \partial_j\phi) + \delta^k_j \delta^{il} \partial_l \phi \right]  T^j_k \, .
      \label{eq:euler1}
\end{align}
The continuity and Euler equations could also be found from energy-momentum conservation, i.e.\ $\nabla_\mu T^\mu_\nu = 0$.

\paragraph{The second moment: Pressure and anisotropic stress}\mbox{}\\
The second moment equation is found by multiplying the Boltzmann equation with the factor $P_i P^j  / P^0$ and then integrating, i.e.\ from the equation 
\begin{equation}
	\sqrt{-g} \int \mathrm{d}^3 P\mspace{1.5mu} \frac{P_i P^j}{P_0 P^0} \left(P^\mu \frac{\partial f}{\partial x^\mu} - P^\mu P^\lambda \Gamma^k_{\mu\lambda}\frac{\partial f}{\partial P^k}\right) = 0 \, .
\end{equation}
Defining the third cumulant as 
\begin{equation}
	\Pi^{\mu\lambda}_\nu \equiv \sqrt{-g}\int \mathrm{d}^3P\mspace{1.5mu} f \frac{P_\nu P^\mu P^\lambda}{P_0 P^0}\,,\quad T^\mu_\nu \equiv \Pi^{\mu 0}_\nu \,,
\end{equation}
and likewise for higher moments, and using the covariant derivative
\begin{equation}
\nabla_\mu \Pi^{j \mu}_i =\partial_\mu \Pi^{j \mu}_i+ \Gamma^j_{\mu\nu} \Pi^{\nu\mu}_i +  \Gamma^{\mu}_{\mu\nu} \Pi^{j\nu}_i  -  \Gamma^\nu_{i\mu} \Pi^{j\mu}_\nu \,,
\end{equation}
one can arrive at the second-moment equation which gives the time-derivative of $T^i_j$
\begin{equation}
     \nabla_\mu \Pi^{j \mu}_i = \Gamma^0_{\mu\nu} \Pi^{j\mu\nu}_i\,,
     \label{eq:second_order1}
\end{equation}
or equivalently
\begin{equation}
     \nabla_0 T^{j}_i = \Gamma^0_{00} T^{j}_i+2\Gamma^0_{0k} \Pi^{jk}_i+\Gamma^0_{kl} \Pi^{jkl}_i - \nabla_k \Pi^{jk}_i\,.
\end{equation}

The time-evolution of $T^j_i$ will depend on the third moment and a contracted version of the fourth moment of the distribution function. This is analogous to the continuity equation where $\dot\delta$ depends on $T^i_0$ and $\delta T^i_i$. 

\subsection{Closing the hierarchy}\label{subsec:closing_the_hierarchy}
The equation for $\nabla_0 T^{j}_i$ is complex and depends on several higher order moments. Instead of solving it directly to get the inputs to the continuity ($\delta T^i_i \sim \delta P$) and Euler ($T^i_j \sim \sigma^i_j, ~\delta P$) equations we will estimate these non-linear terms from their linear counterparts as follows. 

Assuming the ratio $\delta P / \delta\rho$ to be independent of the amplitude of the perturbations, we find
\begin{equation}
	\delta P_{\rm NL}(\vec{k}) \simeq \delta \rho_{\rm NL}(\vec{k}) \left(\frac{\delta P(k)}{\delta\rho(k)}\right)_{\rm L}\,,
\end{equation}
where `L' stands for linear and `NL' for non-linear.

Since $\sigma^i_j$ has the same velocity order as $\delta P$ we will likewise estimate
\begin{equation}
	\sigma^i_{j,\rm NL}(\vec{k})  \simeq \delta \rho_{\rm NL}(\vec{k}) \left(\frac{\sigma^i_j(k)}{\delta\rho(k)}\right)_{\rm L}\,.
\end{equation}

We shall refer the reader to appendix~\ref{section:fourier} for more details on the realisations.

\subsection{The continuity and Euler equations in conservation form}\label{subsec:conservation_form}
For the numerical implementation it is preferable to express the fluid equations using conserved quantities only. To this end, we define the `conserved' density $\varrho$, current $J^i$, pressure $\mathcal{P}$ and anisotropic stress $\varsigma^i_j$ as
\begin{align}
	\varrho &\equiv a^{3(1 + \mathscr{w})}\rho\,, \\
	J^i &\equiv a^4(\rho + P)u^i\,, \\
	\mathcal{P} &\equiv a^{3(1+\mathscr{w})}P\,, \\
	\varsigma^i_j &\equiv (\varrho + \mathcal{P})\sigma^i_j\,,
\end{align}
where the effective equation of state $\mathscr{w}$ is given by
\begin{equation}
	\mathscr{w}(a) \equiv \frac{1}{\ln a}\int_1^a \frac{w(a')}{a'}\, \mathrm{d}a' \,.
\end{equation}
With these variables, the continuity~\eqref{eq:continuity1} and Euler~\eqref{eq:euler1} equations become
\begin{align}
	\dot{\varrho} =& -a^{3\mathscr{w}-1}\partial_i J^i \nonumber \\
	&+ 3aH(w\varrho - \mathcal{P}) \nonumber \\
	&+ a^{3\mathscr{w}-1}J^i\partial_i(3\phi - \psi) \nonumber \\
	&+ 3(\varrho + \mathcal{P})\dot{\phi} \label{eq:continuity_numerical}
\end{align}
and
\begin{align}
	\dot{J}^i =&-\partial^j\biggl[a^{3\mathscr{w} - 1}\frac{J^iJ_j}{\varrho + \mathcal{P}} + a^{-3\mathscr{w}+1}\varsigma^i_j\biggr] \nonumber\\
    &- a^{-3\mathscr{w} + 1}\partial^i\mathcal{P} \nonumber\\
    &- a^{-3\mathscr{w} + 1}\bigl(\varrho + \mathcal{P}\bigr)\partial^i\psi \nonumber\\
    &- a^{3\mathscr{w} - 1} \frac{J^jJ_i}{\varrho+\mathcal{P}}\partial^i\phi  \nonumber\\
    &+ \biggl[a^{3\mathscr{w} - 1}\frac{J^iJ_j}{\varrho + \mathcal{P}} + a^{-3\mathscr{w} + 1}\varsigma^i_j\biggr]\partial^j(3\phi - \psi)  \nonumber\\
    &+ J^i(\dot{\psi} + 5\dot{\phi})\,. \label{eq:euler_numerical}
\end{align}
The claim of conservation of the chosen variables can be checked by spatially averaging \eqref{eq:continuity_numerical} and \eqref{eq:euler_numerical}, indeed leading to $\dot{\bar{\varrho}}=\dot{\bar{J}}^i = 0$.

In our simulations we neglect the difference between $\phi$ and $\psi$ (which is sourced by anisotropic stress), neglect time-derivatives of $\phi$ and $\psi$ and disregard terms of order $\partial_i \phi u^i$ and higher, from which terms with $\partial_i\phi J^i$ and $\partial_i\phi \sigma^i_j$ vanish. Thus only the first two terms of \eqref{eq:continuity_numerical} and the first three terms of \eqref{eq:euler_numerical} are kept.

%%%%%%%%%%%%%%%%%%%%%%%%%%%%%%%%%%%%%%%%%%%%%%%%%%%%%%%%%%%%%%%%%%%%%%%%%%

\section{The linear computation}\label{sec:the_linear_computation}
We compute the linear evolution of all species using the Einstein-Boltzmann code \CLASS{} \cite{Blas:2011rf,Lesgourgues:2011rg}. In the notation of \cite{Ma:1995ey}, the distribution function is expanded as
\begin{equation}
f(\tau, \vec{x}, \vec{p}) = f_0(q)\left[1+ \Psi(\tau,\vec{x}, \vec{p})\right]\,,
\end{equation}
and the evolution equation for $\Psi(q,k,\hat{\vec{q}} \cdot \hat{\vec{k}})$ is then solved in Fourier space. The angular dependence of $\Psi$ is expanded in Legendre multipoles resulting in an infinite hierarchy which is then truncated at some finite $l_\text{max}$. We refer the reader to \cite{Ma:1995ey} for the derivation of the equations.

In this work we do not investigate effects due to the precise neutrino mass hierarchy chosen, and so each neutrino species gets assigned exactly $1/3$ of this total mass $\sum m_\nu$. Though unrealistic, this is a perfectly good choice for method testing as its leaves $\sum m_\nu$ as the only free neutrino parameter, which is precisely the parameter on which the dampening of the matter power spectrum is sensitive.

Neither \CLASS{} nor \CAMB{}~\cite{Lewis:1999bs,Lewis:2002ah} produce accurate neutrino transfer functions at their default precision settings since both codes are optimised to produce accurate CMB and matter power spectra, which do not depend strongly on the late-time neutrino evolution. By increasing the precision parameters of both codes, we have found that agreement can be established at the $1\%$-level or better. We refer the interested reader to appendix~\ref{sec:CLASSCAMB} for more details.

\subsection{$\delta P/\delta \rho$ and $\sigma$ in linear theory}
%%%%%%%%%%%%%%%%%%%%%%%%%%%%%%%%%%%%%%%%%%%%%%%%%%%%%%%%%%%%%%%%%%%%%%%%%%%%%%%%%%%%%%%%%%%%%%%
\begin{table}[tb]
    \begin{center} 
        \begin{tabular}{l c c c c} 
            \hline
            Parameter & $\Lambda$CDM  & $\sum m_\nu = 0.15\,\text{eV}$ & $\sum m_\nu = 0.3\,\text{eV}$ & $\sum m_\nu = 1.2\,\text{eV}$ \\
            \hline
            $A_\text{s}$  & $2.3 \times 10^{-9}$& $2.3 \times 10^{-9}$& $2.3 \times 10^{-9}$& $2.3 \times 10^{-9}$ \\
            $n_\text{s}$ & $1.0$& $1.0$& $1.0$& $1.0$ \\
            $\tau_\text{reio}$ & $0.097765$ & $0.097765$ & $0.097765$ & $0.097765$ \\
            $\Omega_{\text{b}}$ & $0.05$  & 0.05 & 0.05 & 0.05 \\
            $\Omega_{\text{cdm}}$ & $0.250$  & $0.247$ & $0.243$ & $0.224$ \\
            $\Omega_{\nu}$ & $3.48\times 10^{-5}$ & $3.29\times 10^{-3}$ & $6.57\times 10^{-3}$ & $2.61\times 10^{-2}$ \\
            \hline
            $N_{q,\nu}$ & $\cdots$ & 2310 & 1154 & 344 \\  
            $l_{{\rm max},\nu}$ & $\cdots$ & 2000 & 2000 & 1601 \\  
            \hline						
        \end{tabular}
    \end{center}
    \caption{Cosmological parameters and numerical settings for the \CLASS{} runs used.} 
    \label{table:class_parameters} 
\end{table}
%%%%%%%%%%%%%%%%%%%%%%%%%%%%%%%%%%%%%%%%%%%%%%%%%%%%%%%%%%%%%%%%%%%%%%%%%%%%%%%%%%%%%%%%%%%%%%%
%%%%%%%%%
\begin{figure}[t]
\begin{center}
\includegraphics[width=\textwidth]{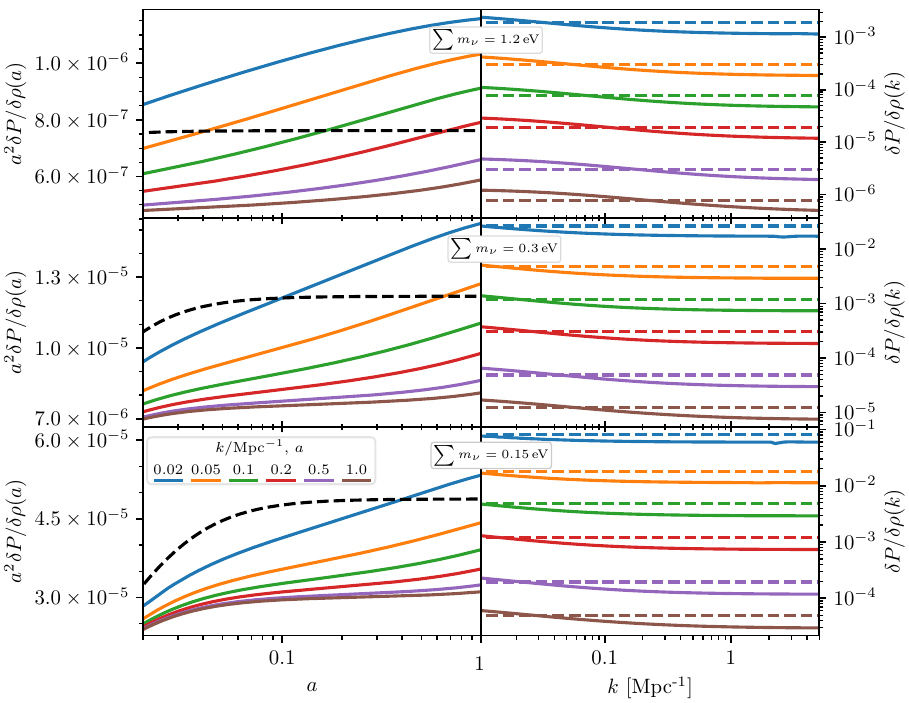}
\end{center}
\caption{Linear $\delta P/\delta\rho$ for $\sum m_\nu = 1.2$ (top), 0.3 (middle) and 0.15 eV (bottom), as function of $a$ (left) and of $k$ (right). Note that all plots share the same legend, so that e.g.\ the blue line on the left corresponds to $k=0.02\,\mathrm{Mpc}^{-1}$ while the blue line on the right corresponds to $a=0.02$. The dashed lines show the corresponding $w$ ($a^2w$ for the left plots). For the \CLASS{} precision settings used, see table~\ref{table:class_parameters}.
\label{fig:dPdrho}}
\end{figure}
%%%%%%%%%

%%%%%%%%%
\begin{figure}[t]
\begin{center}
\includegraphics[width=\textwidth]{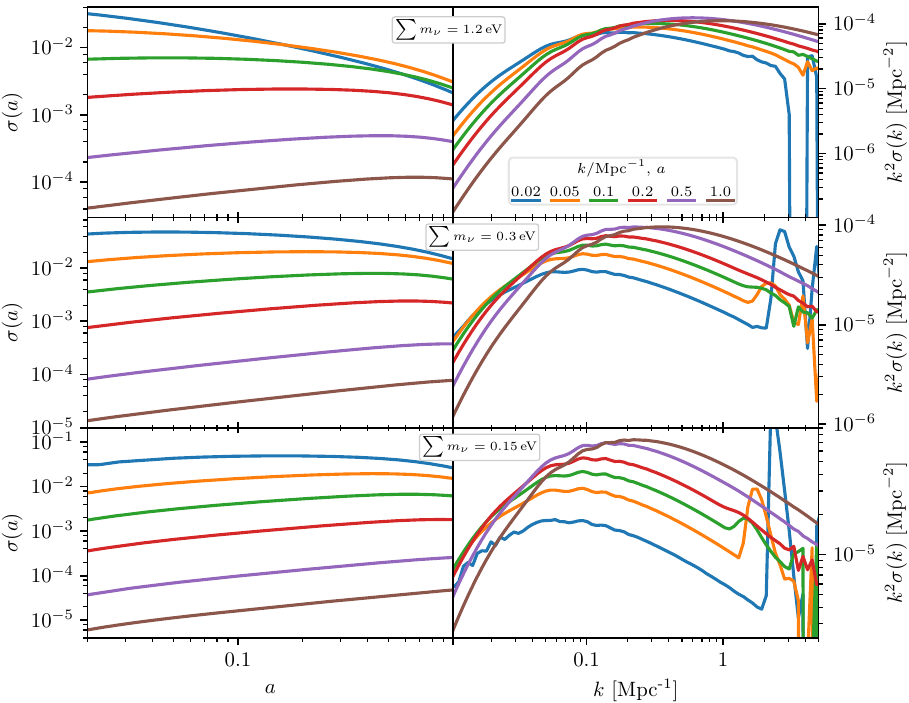}
\end{center}
\caption{Linear $\sigma$ for $\sum m_\nu = 1.2$ (top), 0.3 (middle) and 0.15 eV (bottom), as function of $a$ (left) and of $k$ (right). As in figure~\ref{fig:dPdrho}, all plots share the same legend. For the \CLASS{} precision settings used, see table~\ref{table:class_parameters}.
\label{fig:sigma}}
\end{figure}
%%%%%%%%%

Following the notation of~\cite{Lesgourgues:2011rh}, $\delta P$, $\delta \rho$ and $\sigma$ are related to the Legendre multipoles of $\Psi_l$ as
\begin{align}
\delta \rho &=  4 \pi \left(\frac{T_{\nu,0}}{a}\right)^4 \int_0^\infty f_0 \mspace{1.5mu}\mathrm{d}q\mspace{1.5mu} q^2 \epsilon \Psi_0\,, \\
\delta P &= 4 \pi \left(\frac{T_{\nu,0}}{a}\right)^4 \int_0^\infty f_0 \mspace{1.5mu}\mathrm{d}q\mspace{1.5mu} \frac{q^4}{3 \epsilon} \Psi_0\,, \\
(\bar\rho + \bar P) \sigma &= 8\pi \left(\frac{T_{\nu,0}}{a}\right)^4 \int_0^\infty f_0 \mspace{1.5mu}\mathrm{d}q\mspace{1.5mu} \frac{q^4}{3 \epsilon} \Psi_2\,.
\end{align}
The effective sound speed squared, $\delta P/\delta \rho$, turns out to be extremely challenging to compute numerically. The real $\delta P/\delta \rho$ is expected to be a smooth and monotonic function of $a$, but because the monopole perturbation $\Psi_0$ is highly oscillatory until $q \ll \epsilon$, the discretisation of the systems can easily lead to pathological behaviour of $\delta P/\delta \rho$ unless an extremely large number of momentum bins is used. We find that in some cases it is necessary to use well over 2000 bins (compared to the standard \CLASS{} setting using 5 bins) before the results converge over the required range of $a$.

Furthermore, reflections from the large-$l$ boundary used to close the system of equations can lead to spurious effects such as coherent oscillations of $\delta P/\delta \rho$. We find that we need $l_{\rm max} \gtrsim 2000$ to ensure convergence (compared to $l_{\rm max} = 17$ in the standard setting).
Running \CLASS{} at such extreme precision-settings requires several hundred CPU-hours for a single model, which should be compared to the $\sim 10$ CPU seconds required at default precision. The runtime of \CLASS{} is still a small fraction of the total runtime, but we nevertheless store the \CLASS{} runs to the disk in order to avoid unnecessary recalculations.

Figure~\ref{fig:dPdrho} shows the linear $\delta P/\delta\rho$ resulting from three \CLASS{} computations with different neutrino masses. These are the linear $\delta P/\delta\rho$ values used in our simulations. The slight $k$-dependence makes the local pressure substantially lower than what would be found by approximating $\delta P/\delta\rho \simeq w$, especially for lower (more realistic) neutrino masses. Failure of taking this $k$-dependence into account leads to a mismatch between the equations solved in \CLASS{} and those solved in \CONCEPT{}, which manifests as spurious generation of oscillations in the neutrino density field.

Figure~\ref{fig:sigma} shows the linear $\sigma$ resulting from the same three \CLASS{} computations as was used for figure~\ref{fig:dPdrho}. These are the linear $\sigma$ values used in our simulations. The high $k$ modes have not quite converged at early times, but since $\sigma$ falls off rapidly with $k$ these oscillatory high $k$ modes should only contribute a very low amount of noise to the real-space $\sigma$.

%%%%%%%%%%%%%%%%%%%%%%%%%%%%%%%%%%%%%%%%%%%%%%%%%%%%%%%%%%%%%%%%%%%%%%%%%%%%%%%%%%%%%%%%%%%%%%%
\section{Implementation details}\label{sec:implementation_details}
The methods developed in this paper has been implemented into the \CONCEPT{} code, a new cosmological code capable of simultaneously evolving $N$-body particles (matter) and fluids (neutrinos), interacting under mutual and self-gravity. We have fully integrated the \CLASS{} code into \CONCEPT{}, including an MPI-parallelised method of calling \CLASS{} from \CONCEPT{}, enabling multi-node \CLASS{} computations. With this integration the \CONCEPT{} code has easy access to the evolution of background and linear variables, from which realisations of particle distributions and fluid fields are made. Such realisations are used both for initial condition generation and to close the Boltzmann hierarchy during the $N$-body simulation.

The code is mostly written in Python. For performance, the code may optionally (and preferably) be compiled to C code via Cython. For further optimisations and to lower the Python/Cython barrier, a custom $\text{Python}\rightarrow \text{Cython}$ transpiler is built in as part of \CONCEPT{}. The code is MPI-parallelised with a fixed spatial domain decomposition, dividing up the simulation box into rectangular boxes of equal volume. Each MPI process is then responsible for what goes on within one such domain.

\subsection{Dynamics}\label{subsec:dynamics}
In \CONCEPT{} the collections of either $N$-body particles or fluid variable grids, which are to be evolved dynamically, are grouped into \emph{components}. Particle components consists of a fixed number of particles $N$, of equal mass $m$, whereas fluid components consists of a fixed set of regular, cubic grids of fixed resolution, one for each scalar fluid variable. In the simulations carried out for this paper, baryons and dark matter are grouped together into one particle component, while the neutrino component consists of grids storing $\varrho(\vec{x})$, $J^i(\vec{x})$, $\mathcal{P}(\vec{x})$ and $\varsigma^i_j(\vec{x})$. Of these fluid variables, only $\varrho$ and $J^i$ are treated as non-linear variables, evolved via \eqref{eq:continuity_numerical} and \eqref{eq:euler_numerical} with terms neglected as described in subsection~\ref{subsec:conservation_form}. Since $J^i$ is a vector quantity, this requires 4 grids. The higher-order variables $\mathcal{P}$ and $\varsigma^i_j$ are not evolved non-linearly, but realised at each time step anew. Storing the full $\varsigma^i_j$ would require 6 additional grids (5 if we took advantage of the tracelessness). However, as $\varsigma^i_j$ is only needed once during each time step (in the Euler equation), a single grid is used to store each of its components in turn.

In each time step, all particles and fluid elements are evolved forward in time by the same amount $\Delta \tau$. A leapfrog time integration scheme is used, in which every other time step is either a `kick' or a `drift' step. In a `kick' step, all source terms in the evolution equations are applied. For particle components, the only source term is that of gravity. For fluid components, a term is considered a source term if it is not a flux (divergence) term of one of the lower-order variables $\varrho$ and $J^i$. Explicitly, the partition of terms into flux and source terms is
\begin{align}
	\dot{\varrho} &= -\overbrace{a^{3\mathscr{w}-1}\partial_i J^i}^{\text{flux term}} + \overbrace{3aH(w\varrho - \mathcal{P})}^{\text{source term}} \,, \label{eq:continuity_flux_source} \\
	\dot{J}^i &=-\underbrace{a^{3\mathscr{w} - 1}\partial^j\frac{J^iJ_j}{\varrho + \mathcal{P}}}_{\text{flux term}} - \underbrace{a^{-3\mathscr{w}+1}\Bigl[\partial^j\varsigma^i_j + \partial^i\mathcal{P} + \bigl(\varrho + \mathcal{P}\bigr)\partial^i\psi\Bigr]}_{\text{source terms}} \,. \label{eq:euler_flux_source}
\end{align}

\paragraph{The MacCormack method}\mbox{}\\
In `drift' steps, particle positions $\vec{x}_i$ are updated according to their momenta, while the fluid grids are evolved according to the flux terms of \eqref{eq:continuity_flux_source} and \eqref{eq:euler_flux_source}. To solve these two coupled equations simultaneously, the simple MacCormack \cite{maccormack} finite difference method is used. This method consists of a predictor followed by a corrector step, here illustrated for the flux term of the continuity equation:
\begin{equation}
\begin{dcases}
	\varrho^\star (\vec{x})= \varrho(\vec{x}) - \sum_{i=1}^3\frac{1}{|\mathrm{\Delta}\vec{x}_i|}\bigl[J^i(\vec{x} + \mathrm{\Delta}\vec{x}_i) - J^i(\vec{x})\bigr]\int_{\tau}^{\tau+\mathrm{\Delta} \tau} a^{3\mathscr{w}-1} \mspace{1.5mu}\mathrm{d} \tau\,, \\	
\raisebox{18.1pt}{$\varrho(\vec{x}) \rightarrow$}
\begin{aligned}
&\frac{1}{2}\bigl[\varrho(\vec{x}) + \varrho^\star(\vec{x})\bigr] \\
- &\frac{1}{2}\sum_{i=1}^3\frac{1}{|\mathrm{\Delta}\vec{x}_i|}\bigl[J^{i\star}(\vec{x}) - J^{i\star}(\vec{x} - \mathrm{\Delta}\vec{x}_i)\bigr]\int_{\tau}^{\tau+\mathrm{\Delta} \tau}\! a^{3\mathscr{w}-1}\mspace{1.5mu} \mathrm{d} \tau\,.
\end{aligned}
\end{dcases} \label{eq:maccormack}
\end{equation}
Here the slopes in $\partial_i J^i$ are approximated by the difference between neighbouring grid points along each dimension. The size of the spatial step $|\Delta \vec{x}_i|=\Delta x$ is then just the grid spacing, which in \CONCEPT{} is the same for all dimensions. In the predictor step, a temporary $\varrho^\star(\vec{x})$ grid is build from ``rightward'' differences of $\varrho(\vec{x})$, whereas in the corrector step $\varrho(\vec{x})$ is updated from ``leftward'' differences of $\varrho^\star(\vec{x})$.\footnote{As we are in 3D, ``rightwards'' might be taken to refer to the $(+1, +1, +1)$ direction. In total, 8 possible directions $(\pm 1, \pm 1, \pm 1)$ exist. To avoid spurious generation of anisotropy, \CONCEPT{} cycles through these 8 directions over a period of 8 time steps.} In the corrector step, $J^{i\star}$ is needed, and so the predictor step needs to have been carried out on both the continuity and Euler equation before the corrector step(s) can be applied. This effectively doubles the amount of memory needed to store the non-linear $\varrho$ and $J^i$ variables. In \eqref{eq:maccormack}, all fluid variables on the right-hand side are implicitly evaluated at the current time $\tau$. One might argue that a more self-consistent treatment of the time-dependent function $a^{3\mathscr{w}-1}$ was to similarly evaluate this at time $\tau$, rather than integrating over the time step interval. The choice of keeping the integral is inspired by \cite{Springel:2005mi}.

For particle components, the leapfrog method ensures that $\vec{x}_i$ and $\vec{p}_i$ are always out-of-sync by $\Delta \tau/2$. That is, instead of letting e.g.\ $\vec{x}_i$ evolve by an amount $\Delta \tau$ ahead of $\vec{p}_i$ and then syncing up the system by letting $\vec{p}_i$ evolve by $\Delta \tau$, we first ensure that $\vec{x}_i$ and $\vec{p}_i$ are out-of-sync by $\Delta \tau/2$, leading them to `leapfrog' past one another at every time step. This then treats $\vec{x}_i$ and $\vec{p}_i$ symmetrically, which is the key to the stability of the leapfrog method \cite{Springel:2005mi}. The predictor and corrector steps in which the MacCormack scheme splits up a single (`drift') time step can be thought of as two half time steps, each of duration $\Delta \tau/2$. For fluids, applying gravity out-of-sync by half a time step then means that the gravitational forces applied are those matching the time right after the predictor and before the corrector step. In this way, though gravity is only applied half as often as flux terms, it is applied fairly with respect to the predictor and corrector step.

The bare MacCormack method as illustrated in \eqref{eq:maccormack} is not positivity-preserving, meaning that it is possible for $\varrho(\vec{x})$ to take on slightly negative values, regardless of the smallness of $\Delta \tau$. This can be prevented using e.g.\ total variation diminishing (TVD) extensions \cite{tvdmaccormack,machalinska2014lax} to the method, where flux limiters are applied in order to diminish discontinuities, effectively smoothing out the fluid. We choose a simpler solution, in which a check for negative densities is inserted after each MacCormack step. If a cell with negative density is found, $\varrho$ and $J^i$ of the surrounding block of 27 cells are smoothed out slightly.\footnote{Here, smoothing is performed on each pair of cells in such a way as to preverse their total energy and momentum, with the amount of smoothing inversely proportional to their squared mutual distance.} This operation is quick and only perturbs the fluid in low density regions, as well as in regions with large density gradients.

\begin{figure}[bt]
    \centering
    \includegraphics[width=\textwidth]{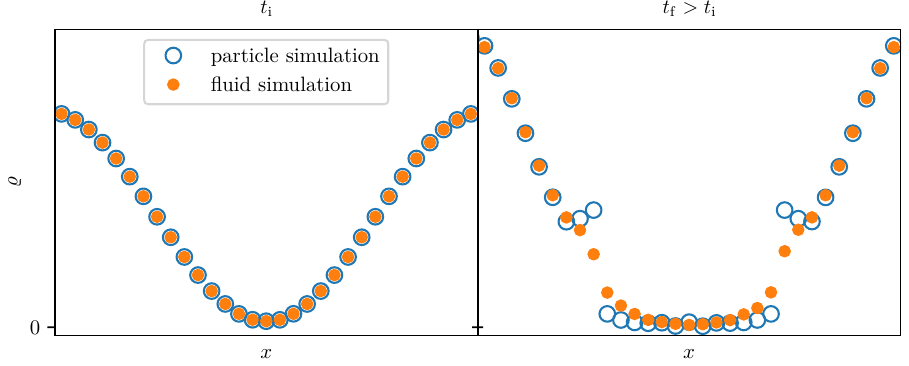}
    \caption{1D test simulations demonstrating the density smoothing extension to the MacCormack method. At time $t_\text{i}$, a sinusoidal wave with minimum density close to 0 is assigned a velocity field which diverges away from this minimum. This wave is then evolved to time $t_\text{f}$ under self-gravity, leading to rapid growth of the low-density region. The same simulation is carried out using particles and a fluid with $w=0$.}
    \label{fig:vacuum_test}
\end{figure}

To stabilise the method we found it necessary to apply the smoothing not only when the density of a fluid cell became negative, but also when a cell lost a large fraction of its total energy. This is needed because the MacCormack method introduces dispersive errors around steep gradients, which makes some kind of artificially added viscosity necessary. Figure~\ref{fig:vacuum_test} demonstrates the effect of this smoothing. We see that the smoothing leaves the overall evolution intact, but erases discontinuities. We do not expect this to seriously perturb our neutrino fluid, as here the pressure term acts as to smooth out the fluid, making it very hard for sharp discontinuities to arise.

\paragraph{Time step size}\mbox{}\\
The size of $\Delta \tau$ is chosen to be as large as possible without affecting the physics. Here the most important limiter is the (global) Courant condition, which we may write as
\begin{equation}
	\Delta \tau < \frac{\Delta x}{\sqrt{3}(c\sqrt{w} + |\vec{u}|_{\text{max}})}\,, \quad |\vec{u}|_{\text{max}} =
	\begin{dcases}
		a^{-1}\max_{i}|\vec{p}_i|/m\,, & \text{(particles)} \\
		a^{3\mathscr{w} - 1}\max_{\vec{x}}\bigl|J^i(\vec{x})/[\varrho(\vec{x}) + \mathcal{P}(\vec{x})]\bigr| \,. & \text{(fluids)}
	\end{dcases} \label{eq:timestep_size}
\end{equation}
Thus the particle or fluid element with largest Courant number sets the global pace of time. For fluid components, \eqref{eq:timestep_size} states that $\Delta \tau$ should be small enough so that a sound wave travelling with speed $c\sqrt{w}$ on top of the bulk flow with maximal peculiar speed $|\vec{u}|_{\text{max}}$ cannot traverse an entire (comoving) grid cell within a single time step.\footnote{As seen from figure~\ref{fig:dPdrho}, the actual, local sound speed $\sqrt{\delta \mathcal{P}(\vec{x})/\delta \varrho(\vec{x})}$ might be somewhat greater than $c\sqrt{w}$, and so \eqref{eq:timestep_size} might not be a strong enough condition. We correct for this by simply multiplying the right-hand side of \eqref{eq:timestep_size} with a small fraction.} For particles, the story is very much similar, except $w\equiv 0$ and the value of $\Delta x$ has to be redefined. As particles by definition do not live on a grid, no inherent grid spacing exists. However, since gravitational interactions are implemented using the particle-mesh method (see subsection~\ref{subsec:gravity}), we use the grid spacing of this mesh as the corresponding $\Delta x$ for particles.

\subsection{Gravity}\label{subsec:gravity}
The \CONCEPT{} code has a rather modular interaction framework, in which different interactions and numerical methods may be assigned to different components. Currently only gravity is implemented, but several methods are available. For particle components, gravity can be computed using either the particle-particle (PP), particle-mesh (PM) or particle-particle-particle-mesh ($\text{P}^3\text{M}$) method. Because both the PP and $\text{P}^3\text{M}$ methods are based on direct summation, they are virtually exact. However, in the current \CONCEPT{} version
PP and P$^3$M are too time-consuming for large simulations and we therefore use the code in PM setup. As this method works by solving the Poisson equation on a mesh, it has an intrinsic limit to the force resolution. 

As fluid components already have an intrinsic resolution limit, nothing is lost by using the PM method, and so only this method is implemented for fluid components. The remainder of this subsection lays out how the PM method is implemented, first for particle-only simulations and then for simulations with both particle and fluid components.

\paragraph{The PM method}\mbox{}\\
The basic strategy of the PM method for particle-only simulations is as follows. Construct the total density field $\rho(\vec{x})$ on a mesh via interpolation. Now Fourier transform this mesh in-place; $\rho(\vec{x})\rightarrow \widetilde{\rho}(\vec{k})$. Convert the grid values to that of the Fourier transformed Newtonian peculiar gravitational potential, $\widetilde{\rho}(\vec{k}) \rightarrow \widetilde{\varphi}(\vec{k})$, using the Poisson equation
\begin{equation}
	\widetilde{\varphi}(\vec{k}) = -\frac{4\pi G a^2}{|\vec{k}|^2}\widetilde{\rho}(\vec{k})\,, \quad \widetilde{\varphi}(\vec{0}) = 0\,. \label{eq:Poisson_Fourier}
\end{equation}
Now perform an inverse Fourier transform to obtain the potential in real space, $\widetilde{\varphi}(\vec{k}) \rightarrow \varphi(\vec{x})$. Finally, use finite difference techniques to obtain approximations for $-\partial_i\varphi$ at each grid point and interpolate the resulting forces back to the particle positions and apply them. Note that a separate mesh is needed to store the forces $-\partial_i\varphi$. The fast Fourier transforms used automatically impose the needed periodic boundary conditions.\footnote{\CONCEPT{} uses the \textsc{fftw} library for MPI-parallel 3D in-place real FFT's.}

Though the same grid in memory is used to store $\rho(\vec{x})$, $\widetilde{\rho}(\vec{k})$, $\widetilde{\varphi}(\vec{k})$ and $\varphi(\vec{x})$ values, we shall refer to this grid consistently as the $\varphi$ grid. In \CONCEPT{}, the cloud-in-cell (CIC) method is used for the interpolations to and from the $\varphi$ grid. This method distributes each of the particles throughout the $\varphi$ grid, with a weight at each grid point $\vec{x}_{\text{m}}$ given by the geometric overlap between the particle and the grid point, where both the particle and the grid point are imagined to have a cubic shape with side lengths equal to the grid spacing of the $\varphi$ mesh, $\Delta x_\varphi$. Denoting the weight at mesh point $\vec{x}_{\text{m}}$ of a particle at $\vec{x}_{\text{p}}$ by $W(\vec{x}_{\text{m}} - \vec{x}_{\text{p}})$, we have
\begin{equation}
	W(x, y, z) = \begin{dcases}
		\biggl(1 - \frac{|x|}{\Delta x_\varphi}\biggr)\biggl(1 - \frac{|y|}{\Delta x_\varphi}\biggr)\biggl(1 - \frac{|z|}{\Delta x_\varphi}\biggr) & \text{if all } |x|, |y|, |z| < \Delta x_\varphi, \\
		0 & \text{otherwise}\,.
	\end{dcases} \label{eq:CIC_weight}
\end{equation}
Note that the weights \eqref{eq:CIC_weight} are only non-zero for the 8 grid points closest to the particle. With the CIC weights \eqref{eq:CIC_weight}, we can write down the interpolation as a convolution,
\begin{equation}
	\rho_{\text{m}}(\vec{x}) = \frac{W(\vec{x})}{\Delta x_\varphi^3}\ast \rho(\vec{x})\,,\quad
	\rho(\vec{x}) = \frac{m}{a^3}\sum_{i=1}^N\delta(\vec{x} - \vec{x}_{i})\,,  \label{eq:particle_interpolation}
\end{equation}
where $\rho_{\text{m}}$ is the mesh-interpolated density and the factor $a^{-3}$ is needed since $\rho$ is physical while $\vec{x}$ is comoving. It is to be understood that numerically, all fields are only defined at the grid points. Thus, though $\vec{x}$ in \eqref{eq:particle_interpolation} is in some sense a discrete variable, this notion clashes with the convolution operation. Instead, let $\vec{x}$ be continuous but make the distinction that numerical grids (e.g.\ $\rho_{\text{m}}$) are only defined at the grid points, unlike their physical counterparts (e.g.\ $\rho$).

In the Poisson equation~\eqref{eq:Poisson_Fourier}, the actual density $\rho$ is needed. Though what lives on the grid is really the interpolated values $\rho_{\text{m}}(\vec{x})\propto W(\vec{x})\ast \rho(\vec{x})$. Simply ignoring this difference leads to errors on scales comparable to the grid spacing $\Delta x_\varphi$. We correct for this by ``undoing'' the CIC convolution while in Fourier space, where the convolution with $W(\vec{x})$ turns into multiplication of $\widetilde{W}(\vec{k})$. The Poisson equation~\eqref{eq:Poisson_Fourier} then becomes
\begin{equation}
	\widetilde{\varphi}_{\text{m}}(\vec{k}) = -\frac{4\pi G a^2}{|\vec{k}|^2} \frac{\Delta x_\varphi^3}{\widetilde{W}(\vec{k})}\widetilde{\rho}_{\text{m}}(\vec{k}) \,, \quad \widetilde{\varphi}_{\text{m}}(\vec{0}) = 0\,, \label{eq:Poisson_Fourier_particles}
\end{equation}
where again, the subscript `m' indicates that this is a numerical grid. The Fourier transform of the CIC weight \eqref{eq:CIC_weight} is
\begin{equation}
	\widetilde{W}(k_x, k_y, k_z) = \Delta x_\varphi^3\biggl[\sinc\biggl(\frac{\Delta x_\varphi k_x}{2}\biggr)\sinc\biggl(\frac{\Delta x_\varphi k_y}{2}\biggr)\sinc\biggl(\frac{\Delta x_\varphi k_z}{2}\biggr)\biggr]^2 \,.
\end{equation}
Equation~\eqref{eq:Poisson_Fourier_particles} results in a properly deconvolved potential on the $\varphi$ grid, though our interest is really the resulting forces at the locations of the particles. Since another CIC interpolation is used to interpolate the forces from the grid points and onto the particles, a total of two CIC deconvolutions are actually needed. The potential actually calculated in particle-only simulations is then $\widetilde{\varphi}_{\text{m}}(\vec{k})\Delta x_\varphi^3/\widetilde{W}(\vec{k})$. Note that this is not the most accurate grid representation of the potential, but it does lead to the most accurate forces after one additional CIC convolution. Finally, just as we replaced $a^{3\mathscr{w}-1}\Delta \tau$ with the integral of $a^{3\mathscr{w}-1}$ over the time step in \eqref{eq:maccormack}, here we replace the $a^2$ in \eqref{eq:Poisson_Fourier_particles} with its average value over the time step. Thus what is really computed on the $\varphi$ grid is
\begin{equation}
	\frac{\Delta x_\varphi^3}{\widetilde{W}(\vec{k})}\widetilde{\varphi}_{\text{m}}(\vec{k}) = -\frac{4\pi G}{|\vec{k}|^2} \biggl[\frac{\Delta x_\varphi^3}{\widetilde{W}(\vec{k})}\biggr]^2 \widetilde{\rho}_{\text{m}}(\vec{k}) \Delta \tau^{-1}\int_{\tau}^{t+\Delta \tau}\! a^2 \mspace{1.5mu}\mathrm{d}\tau \,, \label{eq:Poisson_Fourier_particles_final}
\end{equation}
where again, the DC ($\vec{k}=\vec{0}$) mode is to be disregarded.

\paragraph{Generalising the PM method}\label{sec:CIC}\mbox{}\\
We shall now take a closer look at the needed generalisations to the PM method necessary when both particle and fluid components are present in the same simulation.

We can extend the CIC interpolation scheme to fluid components by specifying a coordinate $\vec{x}_{\mathrm{p}}$ for each fluid element, which should be taken to be at the center of each grid cell. Just as with particles, the CIC interpolation treats fluid elements as cubes with side lengths equal to the grid spacing $\Delta x_\varphi$ of the $\varphi$ grid, regardless of the resolution of the fluid grids themselves.\footnote{Interpolating a homogeneous low resolution fluid grid onto a high resolution $\varphi$ grid thus leaves $\varphi$ with a lot of empty cells. A better fluid interpolation would distribute each fluid cell over the corresponding volume in the $\varphi$ grid. As all simulations for this paper use the same grid size for fluid and potential grids, this is of no concern for our results.} If the fluid grid happens to be of the same resolution as the $\varphi$ grid, all grid points coincide and the CIC interpolation \eqref{eq:CIC_weight} reduces to the trivial mapping, i.e.\ the interpolated and the original grids are equal. In effect then, for fluids with a grid size matching that of the $\varphi$ grid, no CIC interpolation is carried out, and so we are actually worse off if we insist on performing the deconvolutions. Thus, to solve gravity properly, separate computations and $\varphi$ grids are needed for particle and fluid components. Essentially, one grid, $\varphi^{\text{particles}}$, solves the Poisson equation as already described, including the deconvolutions, while another grid, $\varphi^{\text{fluids}}$, solves the Poisson equation without the deconvolutions.\footnote{It is unclear whether carrying out CIC deconvolutions improves or worsens the results for fluids with resolutions different from that of the $\varphi$ grid. In \CONCEPT{}, the $\varphi^{\text{fluids}}$ grid is never deconvolved, regardless of the resolution of the fluid grids.} Importantly, both $\varphi^{\text{particles}}$ and $\varphi^{\text{fluids}}$ should account for the total gravitational potential from both particle and fluid components.

With two separate grids for particle and fluid components, we can construct the mesh-interpolated densities of all particle components $\rho_{\text{m}}^{\text{particles}}$ and of all fluid components $\rho_{\text{m}}^{\text{fluids}}$ by generalisation of \eqref{eq:particle_interpolation}:
\begin{align}
	a^2\rho_{\text{m}}^{\text{particles}}(\vec{x}) &= \frac{W(\vec{x})}{\Delta x_\varphi^3}\ast \sum_{\alpha} m_\alpha\sum_{i=1}^{N_\alpha}\delta(\vec{x} - \vec{x}_{\alpha, i}) \Delta \tau^{-1}\int_\tau^{\tau+\Delta \tau} \frac{\mathrm{d}\tau}{a} \,, \label{eq:particle_interpolation_general} \\
	a^2\rho_{\text{m}}^{\text{fluids}}(\vec{x}) &= \sum_{\alpha} \varrho_\alpha(\vec{x}) \Delta \tau^{-1}\int_\tau^{\tau+\Delta \tau} \frac{\mathrm{d}\tau}{a^{3\mathscr{w}_\alpha+1}}\,, \label{eq:fluid_interpolation_general}
\end{align}
with only the particles being convolved. Here, $\alpha$ in equation~\eqref{eq:particle_interpolation_general} runs over all particle components, while $\alpha$ in equation~\eqref{eq:fluid_interpolation_general} runs over all fluid components. Allowing for multiple fluid components with different $\mathscr{w}$ introduces different integrands of the integrals over the time step, which is why these integrals must be moved from the common potential \eqref{eq:Poisson_Fourier_particles_final} and onto the individual densities. As both $\rho_{\text{m}}^{\text{particles}}$ and $\rho_{\text{m}}^{\text{fluids}}$ are numerical grids, they can immediately be used in the Poisson equation for $\varphi_{\text{m}}^{\text{fluids}}$, while a CIC convolution is required for them to be used in the Poisson equation for $\varphi_{\text{m}}^{\text{particles}}$. Accounting for the single CIC convolution of \eqref{eq:particle_interpolation_general}, we end up with
\begin{align}
	\frac{\Delta x_\varphi^3}{\widetilde{W}(\vec{k})} \widetilde{\varphi}_{\text{m}}^{\mspace{1mu}\text{particles}}(\vec{k}) &= -\frac{4\pi G}{|\vec{k}|^2} \biggl[\biggl(\frac{\Delta x_\varphi^3}{\widetilde{W}(\vec{k})}\biggr)^2 a^2\widetilde{\rho}_{\text{m}}^{\mspace{2mu}\text{particles}}(\vec{k}) + \frac{\Delta x_\varphi^3}{\widetilde{W}(\vec{k})} a^2\widetilde{\rho}_{\text{m}}^{\mspace{2mu}\text{fluids}}(\vec{k})\biggr]\,, \label{eq:Poisson_Fourier_particles_general} \\
	\frac{\Delta x_\varphi^3}{\widetilde{W}(\vec{k})} \widetilde{\varphi}_{\text{m}}^{\mspace{1mu}\text{fluids}}(\vec{k}) &= -\frac{4\pi G}{|\vec{k}|^2} \biggl[\frac{\Delta x_\varphi^3}{\widetilde{W}(\vec{k})} a^2\widetilde{\rho}_{\text{m}}^{\mspace{2mu}\text{particles}}(\vec{k}) +  a^2\widetilde{\rho}_{\text{m}}^{\mspace{2mu}\text{fluids}}(\vec{k})\biggr]	\,, \label{eq:Poisson_Fourier_fluid_general}
\end{align}
where, as usual, we ignore the DC modes. The implementation of the two $\varphi$ grids in \CONCEPT{} is as memory efficient as possible. Equation~\eqref{eq:particle_interpolation_general} and \eqref{eq:Poisson_Fourier_particles_general} and their Fourier duals all live on one grid, while \eqref{eq:fluid_interpolation_general} and \eqref{eq:Poisson_Fourier_fluid_general} and their Fourier duals live on another grid. A single additional grid is used for the final forces $-\partial_i\varphi_{\text{m}}^{\text{particles}}$ and $-\partial_i\varphi_{\text{m}}^{\text{fluid}}$.

From \eqref{eq:Poisson_Fourier_particles_general} and \eqref{eq:Poisson_Fourier_fluid_general}, we see that constructing the two versions of the potential from the densities requires 4 FFT's. Realising $\mathcal{P}_\nu(\vec{x})$ and all 6 components of $\varsigma^i_{j,\nu}(\vec{x})$ requires a total of $1+(1+6)=8$ FFT's, where the one additional FFT is is due to the construction of $\widetilde{\rho}_{\nu}(\vec{k})$, needed for the non-linear realisations as described in subsection~\ref{subsec:closing_the_hierarchy}. We choose to keep $\widetilde{\rho}_{\nu}(\vec{k})$ around as a separate grid, which further increase the memory consumption, but saves us from having to recompute $\widetilde{\rho}_{\nu}(\vec{k})$ $1+6$ times. In total, these 12 FFT's per time step take up about half (for simulations with grid size $1200^3$) of the computation time.

\section{Results}\label{sec:results}
%%%%%%%%%%%%%%%%%%%%%%%%%%%%%%%%%%%%%%%%%%%%%%%%%%%%%%%%%%%%%%%%%%%%%%%%%%%%%%%%%%%%%%%%%%%%%%%
%%%%%%%%%
\begin{figure}[t]
\begin{center}
\includegraphics[width=\textwidth]{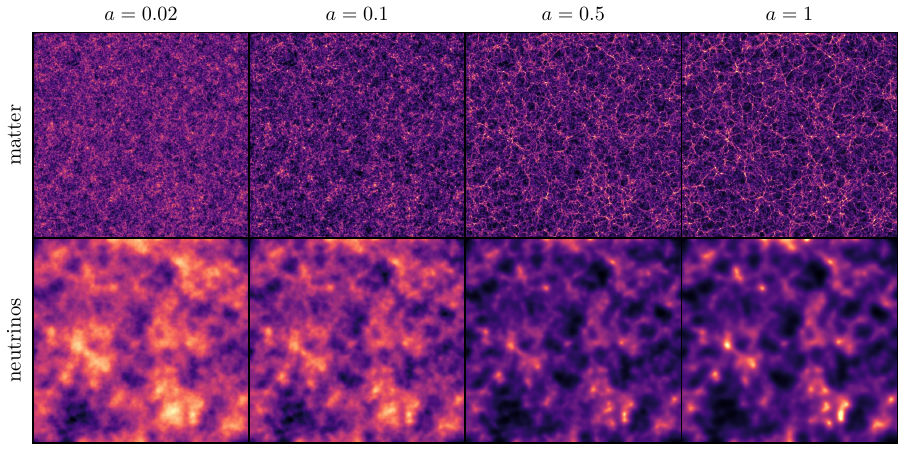}
\end{center}
\caption{Real space density plots for the $\sum m_\nu = 1.2\, \text{eV}$ simulation with grid size $600^3$. All plots have dimensions $512\,\mathrm{Mpc}/h\times 512\,\mathrm{Mpc}/h$ and a depth of $8.5\,\mathrm{Mpc}/h$. For the neutrino plots, the color scalings are linear, though with different absolute values. Each matter plot has its own non-linear color scaling.
\label{fig:slab_renders}}
\end{figure}
%%%%%%%%%

%%%%%%%%%%%%%%%%%%%%%%%%%%%%%%%%%%%%%%%%%%%%%%%%%%%%%%%%%%%%%%%%%%%%%%%%%%%%%%%%%%%%%%%%%%%%%%%
\begin{table}[t]
    \begin{center} 
        \begin{tabular}{l c c c c} 
          Sim &  $\sum m_\nu$ [eV] & $R_{\rm box}~[{\rm Mpc}/h]$ & $N^{\rm part}_{\rm CDM}$ & $N^{\rm grid}_\nu$\\
          \hline
          A&      0.15            & 512       & 600$^3$  & 600$^3$    \\  
          B&      0.15            & 512       & 1200$^3$ & 1200$^3$   \\  
          C&      0.30            & 512       & 600$^3$  & 600$^3$    \\  
          D&      0.30            & 512       & 1200$^3$ & 1200$^3$   \\  
          E&      1.20            & 512       & 600$^3$  & 600$^3$    \\  
          F&      1.20            & 512       & 1200$^3$ & 1200$^3$   \\  
          G&      1.20            & 1024      & 1200$^3$ & 1200$^3$   \\  
          H&      1.20            & 512       & 1680$^3$ & 1680$^3$   \\  
               \hline
      \end{tabular}
      \end{center}
          \caption{\CONCEPT{} simulations used in this work. All simulations are started at $z_{\text{i}}=49$.}
      \label{table:nbody_sims} 
\end{table}

\begin{table}[t]
    \begin{center} 
        \begin{tabular}{l c c c c c} 
           Sim &  $\sum m_\nu$ [eV] & $R_{\rm box}~[{\rm Mpc}/h]$ & $N^{\rm part}_{\rm CDM}$ & $N^{\rm grid}_\nu$ & $N^{\rm part}_{\nu}$\\
          \hline
          X&      0            & 512       &512$^3$  & 0  &0   \\  
          Y&      0.3            & 512       &512$^3$  & 512$^3$  &1024$^3$   \\  

          Z&      1.2            & 512       &512$^3$  & 512$^3$  &1024$^3$   \\  
               \hline
      \end{tabular}
      \end{center}
          \caption{Simulations run with the hybrid neutrino code presented in \cite{Brandbyge:2009ce}.  All simulations are started at $z_{\text{i}}=49$. Neutrinos with $q/T \le 8$ are realised as $N$-body particles at a redshift of $\sim$ 10. The remaining high momentum part is kept as a linear source term on the grid. See \cite{Brandbyge:2009ce} for further information on the hybrid method.} 
      \label{table:nbody_sims_hybrid} 
\end{table}
%%%%%%%%%%%%%%%%%%%%%%%%%%%%%%%%%%%%%%%%%%%%%%%%%%%%%%%%%%%%%%%%%%%%%%%%%%%%%%%%%%%%%%%%%%%%%%%

In order to test the code and compare against the hybrid neutrino method of \cite{Brandbyge:2009ce}, we have performed a suite of different simulations, which are presented in table~\ref{table:nbody_sims} and table~\ref{table:nbody_sims_hybrid}. Table~\ref{table:class_parameters} also specify the $\Omega_\nu$ corresponding to $\sum m_\nu$ in each simulation. We have furthermore used a flat cosmology with $\Omega_\Lambda=0.7$, $\Omega_{\text{b}}=0.05$, $\Omega_{\rm cdm} = 0.25 - \Omega_\nu$, $h=0.7$, $n_{\text{s}}=1$ and $A_{\text{s}} = 2.3\cdot 10^{-9}$.

In order to get the linear theory predictions needed for the initial condition as well as the realisation of $\sigma/\delta \rho$ and $\delta P/\delta\rho$ we have run \CLASS{} with settings given in table~\ref{table:class_parameters}, as described in section~\ref{sec:the_linear_computation}.

Figure~\ref{fig:slab_renders} shows slices of the simulation volume in \CONCEPT{} at 4 different redshifts. The density fields of both the CDM particle component and the neutrino grid component are displayed. The suppression of small scale structure in the neutrino component due to free-streaming is clearly visible.

\subsection{The neutrino power spectra}\label{subsec:neutrino_power_spectra}
\begin{figure}[t]
\begin{center}
\includegraphics[width=\textwidth]{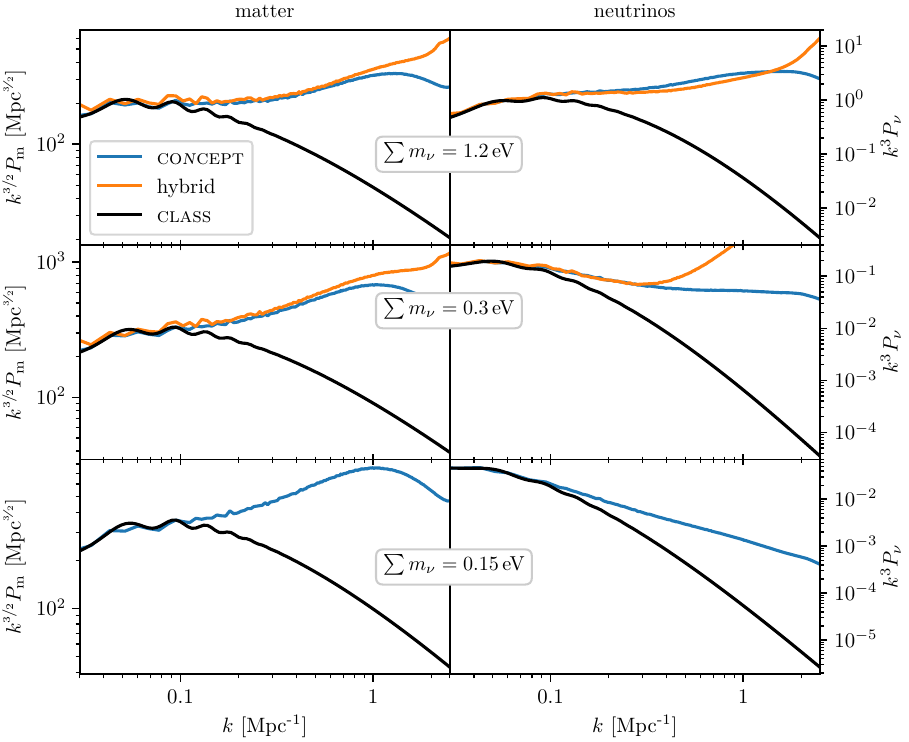}
\end{center}
\caption{Absolute matter (left) and neutrino (right) power spectra at $z=0$ for $\sum m_\nu = 1.2$ eV (top), 0.3 eV (middle) and 0.15 eV (bottom). A significant departure from the linear prediction is seen in the neutrino power for all neutrino masses shown. Excellent agreement between the non-linear predictions of \CONCEPT{} and the hybrid code is achieved out to $k \sim \text{0.2--0.3} \, {\rm Mpc}^{-1}$.
\label{fig:abs}}
\end{figure}

In figure~\ref{fig:abs} we show the absolute matter and neutrino power spectra from simulations~B, D, and F, with parameters given in table~\ref{table:nbody_sims}.

In general we find that there is a very significant increase in neutrino power beyond the linear perturbation theory prediction, in accordance with many previous investigations.
In order to make a more quantitative comparison figure~\ref{fig:abs} also shows matter and neutrino power spectra for the 1.2 eV and 0.3 eV cases, see table~\ref{table:nbody_sims_hybrid}, predicted using the hybrid neutrino grid/particle method from \cite{Brandbyge:2009ce}. 
Out to $k \sim \text{0.2--0.3} \, {\rm Mpc}^{-1}$ the agreement is excellent. At higher $k$ several effects become important, which makes the comparison hard:
\begin{itemize}
\item \CONCEPT{} is run as a particle mesh code and thus has no short range forces included. This leads to lower matter power at late times. 
\item The hybrid code has a significant white noise component in the neutrino power spectrum.
\item For large neutrino masses, the \CONCEPT{} neutrino power spectra develop an unphysical bump for large $k$ at late times. The remainder of this subsection describes the seriousness of this bump and our attempts to understand it.
\end{itemize}
Originally, we hypothesised that the unphysical bump seen in the neutrino power spectra at late times and small scales were due to our simplistic smoothing (described in subsection~\ref{subsec:dynamics}) being inadequate for larger neutrino masses, where a proper smoothing is especially important due to larger spatial gradients in the neutrino fields. We tried replacing the (smoothing extended) MacCormack method with the Kurganov-Tadmor~\cite{kurganov2000} TVD method, but the problem persisted. 

We then speculated that the bump was an artefact originating from our use of the PM method. To test this, we introduced a force-splitting in the gravitational source term for the neutrino species which only kept the (long-range) force up to some cut-off in $k$. The effect was a damping of structure on scales below the cut-off, but the unphysical bump was still present. 
In the end, the cause of the bump remains unknown. It is however not related to how the non-linear system is closed, since the bump is also present when the neutrinos are treated as a perfect fluid inside \CONCEPT{}. This is somewhat visible on the upper-right part of figure~\ref{fig:sigma_importance}.

%%%%%%%%%
%%%%%%%%%

\subsection{Neutrino suppression of the relative total matter power spectra}
%%%%%%%%%
\begin{figure}[t]
\begin{center}
\includegraphics[width=\textwidth]{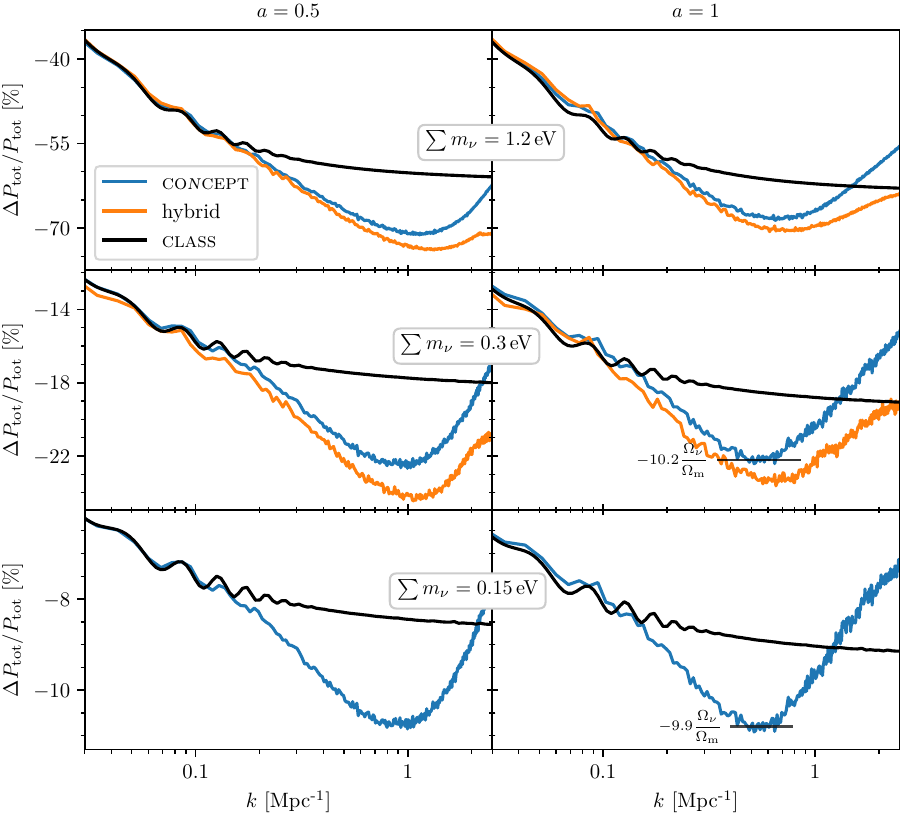}
\end{center}
\caption{The total power spectra (matter and neutrinos) for $\sum m_\nu = 1.2$ (top), 0.3 (middle) and 0.15 eV (bottom), relative to $\Lambda$CDM at $a=0.5$ (left) and $a=1.0$ (right). All panels clearly show a non-linear trough-like shape, the depth of which is known to follow the relation $\sim -10 \Omega_\nu/\Omega_{\rm m}$ for low neutrino masses, to which our findings conform. The non-linear predictions from \CONCEPT{} and the hybrid code agree reasonably well, the largest source of disagreement being the low dynamic range of gravity in \CONCEPT{}.
\label{fig:relative}}
\end{figure}
%%%%%%%%%

In addition to the absolute power spectra it is of interest to investigate the relative suppression of power in models with massive neutrinos relative to standard $\Lambda$CDM with massless neutrinos.
We show these in figure~\ref{fig:relative} for $a=0.5$ and 1 and for the three different choices of $\sum m_\nu$.

For all neutrino masses we see exactly the same trough-like shape of the suppression which was first noticed in
\cite{Brandbyge:2008rv} and which is a generic feature of comparing any model with suppressed structure growth to a standard $\Lambda$CDM model. As non-linear structure formation progresses, larger scales in the neutrino simulations collapse which in turn diminishes the amount of relative suppression and shifts the trough position to smaller $k$-values. These dynamical movements in the relative power spectrum were also found in \cite{Brandbyge:2008js}. 

The maximum suppression in the relative power spectrum at $z=0$ for the lower neutrino masses can be fitted with the relation $\sim -10 \Omega_\nu/\Omega_{\rm m}$, which is in very good agreement with the findings in \cite{Brandbyge:2008rv}. In contrast, the linear theory suppression is roughly given by $\sim -8\Omega_\nu /\Omega_{\rm m}$.

\subsection{Comparison with the hybrid code}\label{subsec:hybrid}
%%%%%%%%%
\begin{figure}[t]
\begin{center}
\includegraphics[width=\textwidth]{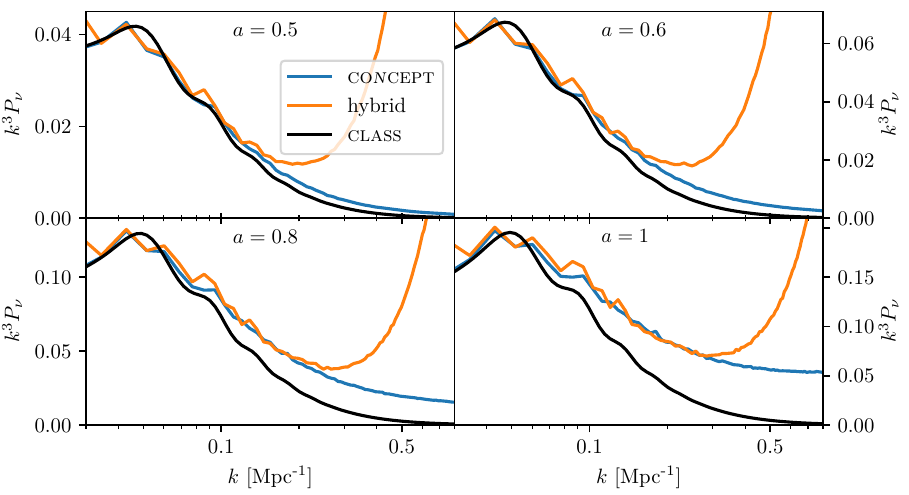}
\end{center}
\caption{Neutrino power spectra for $\sum m_\nu = 0.3$ eV. Non-linearities are seen to build up at late times. For all times excellent agreement between the non-linear predictions of \CONCEPT{} and the hybrid code is seen out to some $k$, after which the hybrid spectrum becomes dominated by white noise.}
\label{fig:nu_compare}
\end{figure}
%%%%%%%%%

Figure~\ref{fig:nu_compare} shows the time evolution of the \CONCEPT{} and hybrid absolute neutrino power spectra for $\sum m_\nu = 0.3$ eV. Excellent agreement is seen for all four times shown, for $k$ below the point of noise domination in the hybrid spectra. This point moves gradually towards smaller scales with time. At $a=1$ we have very good agreement between the \CONCEPT{} and the hybrid neutrino power spectra out to $k\sim \text{0.3} \, {\rm Mpc}^{-1}$.

\subsection{The effect of anisotropic stress}
%%%%%%%%%
\begin{figure}[t]
\begin{center}
\includegraphics[width=\textwidth]{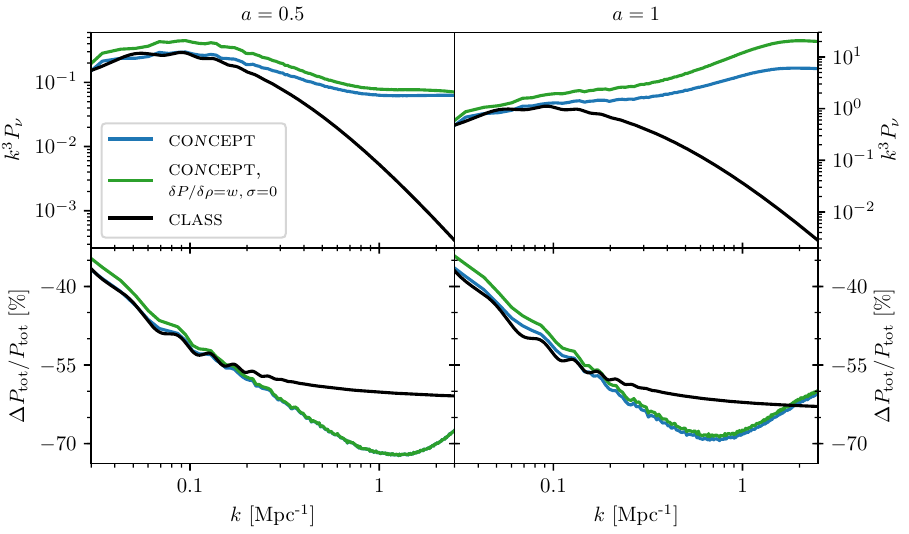}
\end{center}
\caption{Absolute neutrino power spectra (top) and total power spectra (matter and neutrinos) relative to $\Lambda$CDM (bottom), at $a=0.5$ (left) and $a=1.0$ (right), all for $\sum m_\nu = 1.2$ eV. All panels show a full neutrino simulation as well as one where the neutrino component is treated as a perfect fluid, i.e.\ $\delta P/\delta\rho=w$, $\sigma=0$. The two simulations share the exact same initial conditions for both the matter particles and the neutrino fluid, i.e.\ the perfect fluid approximation is switched on at $z=49$. We see that this approximation increases the neutrino power on all scales.
\label{fig:sigma_importance}}
\end{figure}
%%%%%%%%%

In figure~\ref{fig:sigma_importance} we show neutrino power spectra at various $a$ for the 1.2 eV simulation. For comparison we also show the same spectra for a simulation in which we have used $\delta P/\delta \rho=w$ and $\sigma=0$ (i.e.\ the perfect fluid limit). 

From the figure it is evident that this approximation leads to an inaccurate neutrino power spectrum, with too much power on all scales. The effect of neglecting $\sigma$ is known to increase neutrino fluctuations on all sub-horizon scales in linear theory \cite{Hannestad:2004qu}. This effect can be seen in the figure for small $k$, but the figure also demonstrates that a similar effect occurs in the non-linear regime.

\subsection{Convergence and range of validity}
%%%%%%%%%
\begin{figure}[t]
\begin{center}
\includegraphics[width=\textwidth]{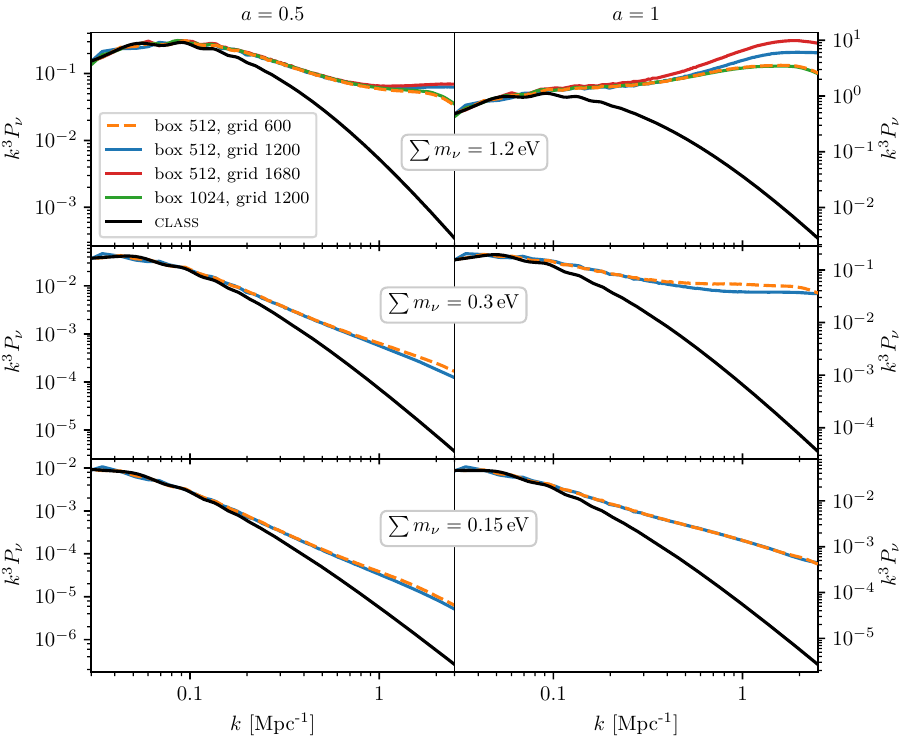}
\end{center}
\caption{Neutrino power spectra for $\sum m_\nu = 1.2$ (top), 0.3 (middle) and 0.15 eV (bottom), at $a=0.5$ (left) and $a=1.0$ (right) for different box and grid sizes. Excellent convergence is seen for the lower neutrino masses, whereas the $\sum m_\nu = 1.2$ neutrino power spectrum fails to converge for increased resolution.
\label{fig:space_convergence}}
\end{figure}
%%%%%%%%%

In figure~\ref{fig:space_convergence} we show neutrino power spectra at $a=0.5$ and $a=1$ for the three different neutrino masses for different choices of box size and grid size.

Several different effects can be seen in this plot. The lowest resolution runs (512 Mpc$/h$ box size and 600$^3$ grid) in general exhibit more power at large $k$ than the other runs. This happens due to errors introduced by the CIC deconvolutions (as described in section~\ref{sec:CIC}), leading to more power in the neutrino component.
It can be seen that this error is very small in the high resolution runs, and excellent convergence is achieved for $0.3\, \text{eV}$ and $0.15\, \text{eV}$. However, for the highest mass ($1.2\, \text{eV}$), the method fails to converge on intermediate scales as discussed in subsection~\ref{subsec:neutrino_power_spectra}.

Even for masses below $0.3\, \text{eV}$, our method will ultimately fail on small enough scales due to the linear closure we employ. If one were to extend our method by one order such that the non-linear stress-tensor was evolved non-linearly, it would be possible to establish the range of scales where the current approach is valid. However, for masses smaller than $0.3\, \text{eV}$, our method match the fully non-linear hybrid code at all scales where the latter can be trusted as shown in subsection~\ref{subsec:hybrid}. We can then establish a very conservative lower bound of $k \lesssim 0.3 \text{Mpc}^{-1}$ for $\sum m_\nu \leq 0.3\, \text{eV}$ at $a=1$.

%However, for the highest mass ($1.2\, \text{eV}$) another effect plays a significant role. As described in sebsection~\ref{subsec:neutrino_power_spectra}, errors arising from the failure of the PM method to resolve the actual gravitational potential leads to an increase in power at intermediate-to-high $k$ and late times. This effect becomes even more significant when the grid size is increased, perhaps simply because of the much greater number of cells on which the error is introduced. The increased neutrino power at $a=1$ seen in the high resolution runs for $1.2\, \text{eV}$ is therefore in all likelihood unphysical. However, for masses more relevant to standard model neutrinos ($\sim 0.15\, \text{eV}$) this effect is absent and therefore even at this stage not a significant source of worry.

\section{Conclusions}\label{sec:conclusions}
%%%%%%%%%%%%%%%%%%%%%%%%%%%%%%%%%%%%%%%%%%%%%%%%%%%%%%%%%%%%%%%%%%%%%%%%%%%%%%%%%%%%%%%%%%%%%%%

We have developed a new method for following neutrinos through non-linear clustering, based on the fully non-linear Boltzmann equation. Our solution is based on the equivalence of the momentum dependent Boltzmann equation to the velocity moment expansion of the same equation. Based on the assumption that moments of order $v^2$ and higher are mainly sourced by linear perturbations we have truncated the velocity hierarchy at this order.
This amounts to solving the fully non-linear continuity and Euler equations for neutrinos, but with $v^2$ source terms derived from the linear perturbation theory solution. 

At the starting point of the simulation these $v^2$ terms can be found simply from the transfer functions provided by \CLASS{} and based on the same set of random numbers used to generate the density and velocity fields.
However, at later times this method fails because terms such as $\delta P$ and $\sigma^i_j$ correlate with the density in the simulation at the given (not the initial) time. We have therefore developed a prescription for how to generate $v^2$ terms correlating with the density field in the simulation and used these as source terms in the continuity and Euler equations.

We have found the method to be both very promising, and established that the method can be used to reliably calculate fully non-linear neutrino power spectra. 
Unlike most methods developed so far which work well for {\it either} small, {\it or} large neutrino masses, the method presented here has the potential to work equally well for both small and intermediate mass neutrinos.
However, we found that for the largest neutrino mass considered (1.2 eV), the current implementation of the method predicts too much structure in the neutrino component in the non-linear regime.

Particle based methods in general suffer from noise related issues when neutrino masses are small because the thermal velocity component requires vast numbers of neutrino particles to sample properly. Conversely, linearised grid based methods break down for high neutrino masses because of non-linearities in the neutrino component.
We have made a comparison between the method developed here and the hybrid method described in \cite{Brandbyge:2009ce} and find that they agree well at scales where the hybrid method yields valid results.

We were unable to use the hybrid method to find a smallest scale below which our method breaks down. To to this we would have to increase the number of neutrino particles in order to reduce the noise. At the present number of neutrino particles, the hybrid simulations are already much more computationally demanding than the corresponding \CONCEPT{} simulations. This then serves to demonstrate the computational superiority of our new method.

The platform for the implementation of neutrinos has been the \CONCEPT{} code~\cite{Dakin:2015uka}, which is made publicly available\footnote{\url{https://github.com/jmd-dk/concept}}. We have fully integrated the code with \CLASS{} such that all linear theory calculations needed by the solver are provided by \CLASS{} without additional input needed from the user.

\section*{Acknowledgements}
We acknowledge computing resources from the Danish Center for Scientific Computing (DCSC). This work was supported by the Villum Foundation.
%%%%%%%%%%%%%%%%%%%%%%%%%%%%%%%%%%%%%%%%%%%%%%%%%%%%%%%%%%%%%%%%%%%%%%%%%%%%%%%%%%%%%%%%%%%%%%%

\appendix
\section{$N$-body realisations of the Boltzmann hierarchy variables}
\label{section:fourier}
We will need to realise several Fourier space transfer functions in real space. The power spectrum of a quantity $Y$ is related to the corresponding transfer function $Y(k)$ from \CLASS{} by
\begin{align}
P_Y(\tau,k) &= 2\pi^2 Y^2(\tau,k) k^{-3} \mathcal{P}_\zeta(k),\\
&= 2\pi^2 A_{\text{s}} Y^2(\tau,k) k^{-3} \( \frac{k}{k_\text{pivot} }\)^{n_\text{s}-1}\, .
\end{align}
Let $\GaussR$ denote a realisation of a Gaussian random field with zero mean such that $\zeta(\vec{k})=\zeta(k) \GaussR(\vec{k})$, where $\zeta$ is the comoving curvature perturbation. The quantity $Y(\tau,\vec{x})$ in real space is then given as
\begin{align}
Y(\tau,\vec{x}) &= \InvFourier{\vec{x}}{\sqrt{P_Y(\tau,k)} \GaussR(\vec{k})} \\
&=\InvFourier{\vec{x}}{ Y(\tau,k) \sqrt{2\pi^2 A_{\text{s}} } k^{-\frac{3}{2}} \( \frac{k}{k_\text{pivot} }\)^{\frac{n_\text{s}-1}{2}}  \GaussR(\vec{k})}.
\end{align}

\subsection{The Zel'dovich approximation}
The transfer function of the Lagrangian displacement field $\zadisp$ for a given species is not directly available in \CLASS{}. However, since $\vec{v} = \dot \zadisp$ the continuity equation for a non-relativistic species in $N$-body gauge \cite{Fidler:2015npa} reads
\begin{equation}
\dot{\delta}(\tau,\vec{x}) = -\nabla \cdot \vec{v}(\tau,\vec{x}) = -\nabla \cdot  \dot{\zadisp}(\tau,\vec{x}) \, .
\end{equation}
Using the boundary condition  $\delta(0,\vec{x}) = 0,\,\zadisp(0,\vec{x}) = \vec{0}$, this equation can be integrated to give
\begin{equation}
\nabla \cdot \zadisp(\tau,\vec{x})  = -\delta(\tau,\vec{x})\, .\label{eq:divdispreal}
\end{equation}
The divergence operator can be easily inverted in Fourier space if we introduce a scalar potential $\Upsilon$ such that $\zadisp(\tau,\vec{x})= \nabla \Upsilon(\tau,\vec{x})$. The equation can then be written as
\begin{equation}
\nabla^2 \Upsilon(\tau,\vec{x}) = -\delta(\tau,\vec{x})\,.
\end{equation}
In Fourier space we have $\ii k^j \Upsilon(\tau,\vec{k}) = \zadispj^j(\tau,\vec{k})$,  which leads to
\begin{align}
-k^2  \Upsilon(\tau,\vec{k}) &= -\delta(\tau,\vec{k}) \\
 \Rightarrow \Upsilon(\tau,\vec{k}) &= \frac{\delta(\tau,\vec{k})}{k^2} \\
  \Rightarrow \zadispj^j(\tau,k)&= \frac{\ii k^j}{k^2} \delta(\tau,k) \,,
\end{align}
where the last equality follows since an equation for quantities in Fourier space also holds for the corresponding transfer functions.
Explicitly, this gives
\begin{equation}
\zadispj^j(\tau,\vec{x}) = \InvFourier{\vec{x}}{ \( \frac{\ii k^j}{k^2} \delta(\tau,k) \) \sqrt{2\pi^2 A_{\text{s}} } k^{-\frac{3}{2}} \( \frac{k}{k_\text{pivot} }\)^{\frac{n_{\text{s}}-1}{2}}\GaussR(\vec{k})}\,.
\end{equation}

\subsection{Density and velocity fields}
The density field can be directly realised from the transfer function $\delta(\tau,k)$:
\begin{equation}
\delta(\tau,\vec{x}) = \InvFourier{\vec{x}}{ \delta(\tau,k) \sqrt{2\pi^2 A_{\text{s}} }  k^{-\frac{3}{2}} \( \frac{k}{k_\text{pivot} }\)^{\frac{n_{\text{s}}-1}{2}}
\GaussR(\vec{k})}\,. \label{eq:delta_realisation}
\end{equation}
Because \CLASS{} solves for the divergence $\theta$ of the velocity field, we must invert the divergence operator in Fourier space like we do for the displacement field. We find
\begin{equation}
v^j(\tau,\vec{x}) = \InvFourier{\vec{x}}{ \( -\frac{\ii k^j}{k^2} \theta(\tau,k) \) \sqrt{2\pi^2 A_{\text{s}} } k^{-\frac{3}{2}} \( \frac{k}{k_\text{pivot} }\)^{\frac{n_{\text{s}}-1}{2}}
\GaussR(\vec{k})}\,, \label{eq:velovity_realisation}
\end{equation}
by the identification $\zadispj^j \mapsto v^j$, $\delta \mapsto -\theta$ in equation~\eqref{eq:divdispreal}. Note that \eqref{eq:velovity_realisation} describes the velocity field of a fluid as well as the velocity field that should be used to set particle velocities when generating particle initial conditions using the Zel'dovich approximation. 

\subsection{Anisotropic stress}
\CLASS{} solves for a quantity $\sigma(\tau,k)$ called the scalar anisotropic stress. Following \cite{Ma:1995ey} we define $\Sigma^i_j$ as the trace-free contribution to the energy-momentum tensor\footnote{The $u^iu_j$ part present in \eqref{eq:EMT} has been left out as we now work in linear theory.},
\begin{equation}\label{eq:Tij}
T^i_j(\tau,\vec{x}) = (\bar P + \delta P) \delta_i^j + \Sigma^i_j(\tau,\vec{x}) \,.
\end{equation} 
As $\Sigma^i_j(\tau,\vec{x})$ is a symmetric, trace-free rank 2 tensor, it has 5 degrees of freedom: 2 tensor, 2 vector and one scalar degree of freedom. We can define the scalar potential $\gamma$ implicitly by
\begin{align}
\Sigma^i_j(\tau,\vec{x}) &= \(\nabla^i \nabla_j - \frac{1}{3} \delta^i_j \nabla^2  \) \gamma(\tau,\vec{x}) \,,
\end{align}
which in Fourier space becomes
\begin{align}
\Sigma^i_j(\tau,\vec{k}) &= -k^2\(\hat{\vec{k}}^i \hat{\vec{k}}_j - \frac{1}{3} \delta^i_j  \) \gamma(\tau,\vec{k})\,.
\end{align}
We can now compare this to the definition of $\sigma$ (equation~22 in \cite{Ma:1995ey}):
\begin{align}
\bar \rho (1+w) \sigma(\tau,\vec{k}) &=  -\(\hat{\vec{k}}^j \hat{\vec{k}}_i - \frac{1}{3} \delta^j_i  \) \Sigma^i_j(\tau,\vec{k}) \label{eq:aniso1}\\
&= k^2 \( \hat{\vec{k}}^j \hat{\vec{k}}_i \hat{\vec{k}}^i \hat{\vec{k}}_j + \frac{1}{9} \delta^j_i   \delta^i_j -\frac{2}{3}\hat{\vec{k}}^j \hat{\vec{k}}_i  \delta^i_j  \)  \gamma(\tau,\vec{k}) \\
&= \frac{2}{3} k^2   \gamma(\tau,\vec{k})\,.
\end{align}
The final expression then becomes 
\begin{equation}
\Sigma^i_j(\tau,\vec{x}) = \InvFourier{\vec{x}}{ -\frac{3}{2} \bar \rho (1+w) \( \hat{\vec{k}}^i \hat{\vec{k}}_j - \frac{1}{3} \delta^i_j \) \sigma(\tau,k) \sqrt{2\pi^2 A_{\text{s}} } k^{-\frac{3}{2}} \( \frac{k}{k_\text{pivot} }\)^{\frac{n_{\text{s}}-1}{2}}\GaussR(\vec{k})}. \label{eq:Sigma_realisation}
\end{equation}
By comparing equation~\eqref{eq:Tij} to the linearised version of equation~\eqref{eq:EMT}, we find $\Sigma^i_j(\tau,\vec{x}) =  \bar \rho (1+w) \sigma^i_j(\tau,\vec{x})$ leading to
\begin{equation}
\sigma^i_j(\tau,\vec{x}) = \InvFourier{\vec{x}}{-\frac{3}{2} \( \hat{\vec{k}}^i \hat{\vec{k}}_j - \frac{1}{3} \delta^i_j \) \sigma(\tau,k) \sqrt{2\pi^2 A_{\text{s}} } k^{-\frac{3}{2}} \( \frac{k}{k_\text{pivot} }\)^{\frac{n_{\text{s}}-1}{2}}\GaussR(\vec{k})}. \label{eq:sigma_relisation}
\end{equation}

\subsection{Non-linear realisations}
Since the linear pressure perturbation $\delta P$ is a scalar, it can be realised in a manner similar to that of $\delta$, \eqref{eq:delta_realisation}:
\begin{align}
	\delta P(\tau,\vec{x}) &= \InvFourier{\vec{x}}{ \delta P(\tau,k) \sqrt{2\pi^2 A_{\text{s}} }  k^{-\frac{3}{2}} \( \frac{k}{k_\text{pivot} }\)^{\frac{n_{\text{s}}-1}{2}}
\GaussR(\vec{k})} \label{eq:deltaP_realisation} \\
	&= \InvFourier{\vec{x}}{\frac{\delta P(\tau, k)}{\delta(\tau, k)}\delta(\tau, \vec{k})}\,, \label{eq:deltaP_realisation2}
\end{align} 
where $\delta(\tau, \vec{k})=\Fourier{\vec{x}}{\delta(\tau,\vec{x})}$ is simply the content of the bracket in \eqref{eq:delta_realisation}. With this interpretation of $\delta(\tau, \vec{k})$, the resulting $\delta P(\tau, \vec{x})$ from \eqref{eq:deltaP_realisation} is purely linear. If we now upgrade $\delta(\tau, \vec{k})$ to be the non-linear density contrast present in the simulation at any time $\tau$, \eqref{eq:deltaP_realisation} yields our estimate of the non-linear $\delta P$ at any time $\tau$. Comparing \eqref{eq:deltaP_realisation} with \eqref{eq:deltaP_realisation2}, we can write
\begin{equation}
\GaussR(\tau, \vec{k}) = \frac{1}{\sqrt{2\pi^2 A_{\text{s}} }}k^{\frac{3}{2}}\biggl(\frac{k}{k_{\text{pivot}}}\biggr)^{\frac{1-n_{\text{s}}}{2}}\frac{\delta(\tau, \vec{k})}{\delta(\tau, k)}\,, \label{eq:R_time_evolved}
\end{equation}
with $\GaussR(\tau, \vec{k})$ being the time evolved random phases, coinciding with $\GaussR(\vec{k})$ at the initialisation time. The time evolution of the estimated non-linear $\delta P$ from \eqref{eq:deltaP_realisation2} are then due to two effects: the time evolution of the linear transfer function of $\delta P$ itself, as well as the non-linear time evolution of the underlying random field $\GaussR$. Simply ignoring this last effect and using the same $\GaussR(\vec{k})$ throughout time leads to a mismatch between the actual and supposed phases, resulting in large errors.

Multiplying both $\delta(\tau, k)$ and $\delta(\tau, \vec{k})$ in \eqref{eq:deltaP_realisation2} by $\bar{\rho}$, we can write the approximation generated by interpreting $\delta(\tau, \vec{k})$ as the non-linear density contrast as
\begin{equation}
	\delta P_{\rm NL}(\tau, \vec{k}) \simeq  \frac{\delta P_{\text{L}}(\tau, k)}{\delta\rho_{\text{L}}(\tau, k)} \delta \rho_{\rm NL}(\vec{k})\,, \label{eq:dP_nonlinear_realisation}
\end{equation}
where `L' and `NL' stands for `linear' and `non-linear', respectively. Thus the approximation corresponds to the assumption that $\delta P / \delta\rho$ (and hence the sound speed) is independent of the amplitude of the perturbations.

As we also want to realise $\sigma^i_j(\tau,\vec{x})$ throughout the simulation timespan, we similarly need an estimate of the non-linear $\sigma^i_j(\tau,\vec{x})$. Comparing \eqref{eq:sigma_relisation} to \eqref{eq:deltaP_realisation}, we see that the only difference is the factor $-3/2\bigl(\hat{\vec{k}}^i \hat{\vec{k}}_j - \delta^i_j/3\bigr)$, and so
\begin{align}
	\sigma^i_{j,\text{NL}}(\tau, \vec{k}) &\simeq -\frac{3}{2}\biggl(\hat{\vec{k}}^i \hat{\vec{k}}_j - \frac{1}{3}\delta^i_j\biggr)\frac{\sigma_{\text{L}}(\tau, k)}{\delta\rho_{\text{L}}(\tau, k)} \delta\rho_{\text{NL}}(\tau, \vec{k}) \,,\label{eq:sigma_nonlinear_realisation} \\
    \sigma^i_{j,\text{NL}}(\tau, \vec{x}) &= \InvFourier{\vec{x}}{\sigma^i_{j,\text{NL}}(\tau, \vec{k})}\,.
\end{align}
Similarly, this approximation corresponds to the assumption that $\sigma/ \delta\rho$ is independent of the amplitude of the perturbations. Since $\sigma^i_j$ has the same velocity order as $\delta P$, the two approximations should be equally valid.

Interpreting \eqref{eq:R_time_evolved} as being solely the evolved phases disregards the fact that on top of the shifting phases we also have the non-linear growth of $\delta(|\vec{k}|)$. Thus, equations~\eqref{eq:dP_nonlinear_realisation} and \eqref{eq:sigma_nonlinear_realisation} do not only supply $\delta P_{\mathrm{NL}}$ and $\sigma^i_{j,\mathrm{NL}}$ with the correctly evolved phases, but inevitably also injects non-linearity. This can be avoided by replacing the linear transfer function $\delta \rho(\tau, k)$ with its non-linear counterpart, namely the square root of the non-linear power spectrum:
\begin{align}
	\delta P_{\rm NL}(\tau, \vec{k}) &\simeq \sqrt{\frac{P_{\delta P_{\mathrm{L}}}(\tau, k)}{P_{\delta\rho_{\mathrm{NL}}}(\tau, k)}} \delta\rho_{\mathrm{NL}}(\tau, \vec{k})\,, \label{eq:dP_nonlinear_realisation_only_phases} \\
	\sigma^i_{j,\text{NL}}(\tau, \vec{k}) &\simeq -\frac{3}{2}\biggl(\hat{\vec{k}}^i \hat{\vec{k}}_j - \frac{1}{3}\delta^i_j\biggr) \sqrt{\frac{P_{\sigma_{\mathrm{L}}}(\tau, k)}{P_{\rho_{\rm NL}}(\tau, k)}} \delta\rho_{\text{NL}}(\tau, \vec{k}) \,, \label{eq:sigma_nonlinear_realisation_only_phases}
\end{align}
where the transfer function of the target variable also has been replaced by the (linear) power spectrum,
\begin{equation}
	P_{Y_{\mathrm{L}}}(\tau, k) = 2\pi^2 A_{\mathrm{s}}k^{-3}\biggl(\frac{k}{k_{\mathrm{pivot}}}\biggr)^{n_{\mathrm{s}}-1} Y^2(\tau, k) \,,
\end{equation}
in order to cancel out the factors otherwise introduced by exchanging a transfer functions for a (square root of a) power spectrum. Equations~\eqref{eq:dP_nonlinear_realisation_only_phases} and \eqref{eq:sigma_nonlinear_realisation_only_phases} \emph{can} then be used in place of \eqref{eq:dP_nonlinear_realisation} and \eqref{eq:sigma_nonlinear_realisation}. This leaves us with two separate realisation schemes with no obvious best choice.

Since what we really realize is $\varsigma^i_j\propto \Sigma^i_j = (\rho + P)\sigma^i_j$, we further have the choice of whether to use $\bar{\rho}(1+w)$ as in \eqref{eq:Sigma_realisation} (inside or outside of the Fourier transform) or the non-linear $\rho+P$ (outside the Fourier transform), as in
\begin{equation}
	\Sigma^i_{j,\text{NL}}(\tau, \vec{k}) \simeq \bigl(\rho_{\rm NL}(\tau, \vec{k}) + P_{\rm NL}(\tau, \vec{k})\bigr)\InvFourier{\vec{x}}{ -\frac{3}{2}\biggl(\hat{\vec{k}}^i \hat{\vec{k}}_j - \frac{1}{3}\delta^i_j\biggr) \frac{\sigma_{\text{L}}(\tau, k)}{\delta\rho_{\text{L}}(\tau, k)} \delta\rho_{\text{NL}}(\tau, \vec{k}) }\,, \label{eq:Sigma_nonlinear_realisation_extra_nonlin}
\end{equation}
again with the further possibility of replacing transfer functions with power spectra.
Once again, these choices come down to whether we wish to further inject non-linearity into $\Sigma^i_j$. We generally achieved better results with this added non-linearity, and all plots in this paper have been produced using \eqref{eq:dP_nonlinear_realisation} and \eqref{eq:Sigma_nonlinear_realisation_extra_nonlin} as non-linear realisation schemes. Further studying of these different realization schemes --- which are all equivalent in linear theory but differ in non-linear theory --- would be very interesting.

\section{Comparison between CLASS and CAMB}\label{sec:CLASSCAMB}
\begin{figure}[t]
\begin{center}
\includegraphics[width=\textwidth]{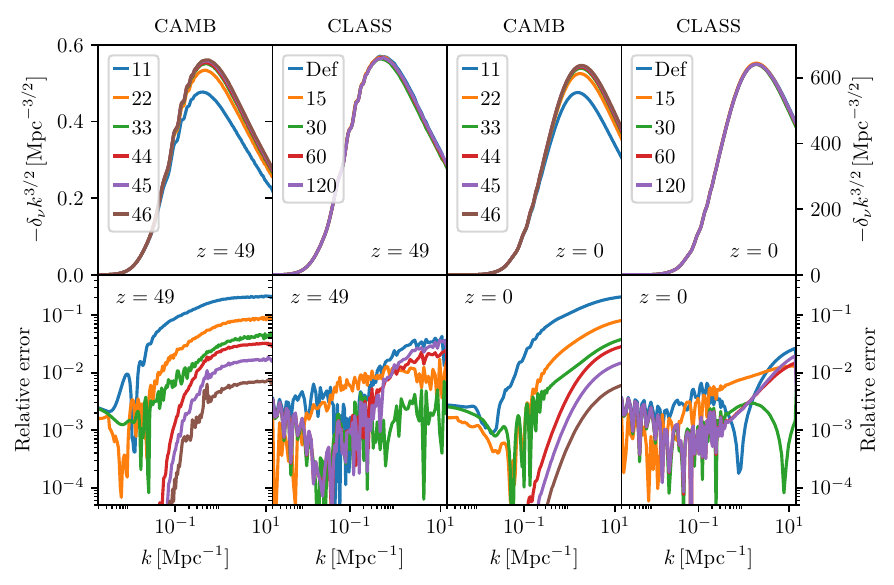}
\end{center}
\caption{The density transfer function $\delta_\nu(k)$ multiplied by $-k^{3/2}$ for $\sum m_\nu = 1.2\,\text{eV}$ at redshift $z=49$ (left) and $z=0$ (right). For \CAMB{}, the first digit corresponds to \textttj{accuracy\_boost} and the second to \textttj{laccuracy\_boost}. For \CLASS{}, the legend denotes the value of both $N_q$ and $l_{\text{max},\nu}$, except for `Def' which denotes the default settings.
\label{fig:CAMBCLASS}}
\end{figure}
Neither \CLASS{} nor \CAMB{} produce accurate neutrino transfer functions at their default precision settings. The reason is simply that these precision settings are tuned for the total matter power spectrum which is usually dominated by the cold matter. The precision of the neutrino evolution is mainly controlled by the number of momentum bins, $N_q$ and the cut-off in the Boltzmann hierarchy $l_{\text{max},\nu}$.

The momentum sampling $q$ in $\CLASS{}$ is automatic as discussed in detail in \cite{Lesgourgues:2011rh}. An optimal choice of quadrature method is found by requiring that the distribution function multiplied by a test-function can be computed at sufficient accuracy. This has the virtue of being independent of the actual distribution function. However, in order to obtain high-precision neutrino transfer functions from \CLASS{}, we had to use the manual quadrature strategy which was introduced in \CLASS{} \textttj{v2.6.2}. We use \textttj{quadrature strategy = 3} which means that \CLASS{} uses a trapezoidal rule on a uniform grid from $0$ to $q_\text{max}$ with $N_q+1$ points. The first point at $q=0$ is not actually evolved since all integrands would anyway vanish.

For the high-precision \CLASS{} runs we turn off the fluid approximation by setting the parameter \textttj{ncdm\_fluid\_approximation = 3}, and this requires us to increase the $l_\text{max}$ cutoff in the Boltzmann hierarchy considerably since the lowest multipoles now have more time to get polluted by the unphysical reflection of power at $l_\text{max}$.

The agreement between  \CLASS{} and \CAMB{} for the neutrino transfer functions has never been checked in any detail, so we have conducted a preliminary convergence test for the case $\sum m_\nu = 1.2~\text{eV}$. As we have shown in figure~\ref{fig:CAMBCLASS}, the two codes can be brought into agreement at the $1\%$-level.

We matched the cosmological model in the two codes and used the precision settings given in table~\ref{table:cambclass}.
\begin{table}[t]
    \begin{center} 
\begin{tabular}{ll}
\CAMB{} & \CLASS{} \\
\hline
\textttj{high\_accuracy\_default = T}  &   \textttj{ncdm\_fluid\_approximation = 3} \\
\textttj{transfer\_high\_precision = T} & \textttj{Quadrature strategy = 3}  \\  
\textttj{massive\_nu\_approx = 0} &  \textttj{Maximum q = 15}   \\
\textttj{accuracy\_boost = 1-4} & \textttj{Number of momentum bins = 15 (30, 60, 120)}  \\
\textttj{l\_accuracy\_boost = 1-7} & \textttj{l\_max\_ncdm = 15 (30, 60, 120)}\\
\hline
\end{tabular}
\end{center}
\caption{Precision settings used in the \CAMB{} and \CLASS{} comparison runs.}
\label{table:cambclass} 
\end{table}
For the \CAMB{}-runs, the first and second digit in the legend of figure~\ref{fig:CAMBCLASS} refers to the value of \textttj{accuracy\_boost} and \textttj{l\_accuracy\_boost}, respectively. For the \CLASS{}-runs the two precision parameters, \textttj{Number of momentum bins} and \textttj{l\_max\_ncdm} were set to the same value and varied together, and the legend refers to the value of both.

We used a $47$ \CAMB{} run (not included in the figure) as the common reference for computing the relative error of both codes. We see that the $11$ setting of \CAMB{} generates a 20\% error. We emphasise that this is not a problem for standard cosmological analyses, but it could be an issue when used as initial condition for neutrino simulations. For the \CLASS{} runs we see that the error seems to increase for large $k$ when the precision is increased beyond $l_{\text{max},\nu}=N_q=30$, which we take as an indication that the $47$ \CAMB{}-run is not yet numerically converged. An agreement at the $1\%$-level that we have established is however enough for the present implementation.

\bibliographystyle{utcaps}
\providecommand{\href}[2]{#2}\begingroup\raggedright\endgroup

\end{document}